\documentclass[aps,prl,twocolumn,superscriptaddress,amsmath,amssymb,floatfix,apsrev,bibnotes]{revtex4-2}

\usepackage{tabularx}
\usepackage{wasysym}
\usepackage{bm}
\usepackage{epsfig,subfigure}
\usepackage{multirow}
\usepackage{graphicx}
\usepackage{xcolor}
\usepackage{amsfonts}
\usepackage{exscale}
\usepackage{amsbsy}
\usepackage{float}
\usepackage{color}
\usepackage{comment}
\usepackage{braket}
\graphicspath{{figs/},{../figs/}}
\usepackage[colorlinks,urlcolor=blue,linkcolor=blue,anchorcolor=blue,citecolor=blue]{hyperref}

\newcommand{\ta}{\tilde{a}}
\newcommand{\tb}{\tilde{b}}
\newcommand{\tn}{\tilde{n}}
\newcommand{\tK}{\tilde{K}}
\newcommand{\tchi}{\tilde{\chi}}

\newcommand{\printauthorlist}{{
  \expandafter\let\csname \textsuperscript \endcsname\@gobble
  \AB@authlist}
}

\begin{document}

\title{Floquet Flux Attachment in Cold Atomic Systems}

\author{Helia Kamal}
\thanks{These authors contributed equally to this work.}
\affiliation{Department of Physics, Harvard University, Cambridge, MA 02138, USA}

\author{Jack Kemp}
\thanks{These authors contributed equally to this work.}
\affiliation{Department of Physics, Harvard University, Cambridge, MA 02138, USA}

\author{Yin-Chen He}
\thanks{These authors contributed equally to this work.}
\affiliation{Perimeter Institute for Theoretical Physics, Waterloo, ON N2L 2Y5, Canada}

\author{Yohei Fuji}
\affiliation{Department of Applied Physics, University of Tokyo, Tokyo 113-8656, Japan}

\author{Monika Aidelsburger}
\affiliation{Max-Planck-Institut f\"ur Quantenoptik, 85748 Garching, Germany}
\affiliation{Faculty of Physics, Ludwig-Maximilians-Universit\"{a}t M\"{u}nchen, Schellingstr. 4, D-80799 Munich, Germany}
\affiliation{Munich Center for Quantum Science and Technology (MCQST), Schellingstr. 4, D-80799 Munich, Germany}

\author{Peter Zoller}
\affiliation{Institute for Theoretical Physics, University of Innsbruck, Innsbruck, 6020, Austria}
\affiliation{Institute for Quantum Optics and Quantum Information of the Austrian Academy of Sciences, Innsbruck, 6020, Austria}

\author{Norman Y.~Yao}
\affiliation{Department of Physics, Harvard University, Cambridge, MA 02138, USA}

\begin{abstract}
Flux attachment provides a powerful conceptual framework for understanding certain forms of topological order, including most notably the fractional quantum Hall effect.  
Despite its ubiquitous use as a theoretical tool, directly realizing flux attachment in a microscopic setting remains an open challenge. Here, we propose a simple approach to realizing flux attachment in a periodically-driven (Floquet) system of either spins or hard-core bosons. We demonstrate that such a system naturally realizes correlated hopping interactions and provides a sharp connection between such interactions and flux attachment. Starting with a simple, nearest-neighbor, free boson model, we find evidence---from both a coupled-wire analysis and large-scale density matrix renormalization group simulations---that Floquet flux attachment stabilizes the bosonic integer quantum Hall state at $1/4$ filling (on a square lattice), and the  Halperin-221 fractional quantum Hall state at $1/6$ filling (on a honeycomb lattice).
At $1/2$ filling on the square lattice, time-reversal symmetry is instead spontaneously broken and bosonic integer quantum Hall states with opposite Hall conductances are degenerate.
Finally, we propose an optical-lattice-based implementation of our model on a square lattice and  discuss prospects for adiabatic preparation as well as effects of Floquet heating.
\end{abstract}

\maketitle

Unlike more conventional states, topological phases cannot be identified by their pattern of symmetry breaking, but rather, by the underlying structure of their entanglement \cite{Chen2010}. 
Despite tremendous recent advances in the theoretical classification \cite{Chen2013, Lu2012} of both intrinsic topological order \cite{Tsui1982, Laughlin1983} and symmetry-protected topological (SPT) order \cite{Chen2013,Haldane1983,Pollmann2010,Vishwanath2013}, there are, as yet, few guiding principles for obtaining simple realizations of strongly-interacting topological phases. 

From a conceptual viewpoint, one such principle, which underlies our understanding of the fractional quantum Hall effect \footnote{As well as putative spin liquid phases of frustrated magnets.}, is the notion of flux attachment \cite{Girvin1987,Zhang1989, Jain1989, Zhang1992}; the conventional picture states that Coulomb repulsion has the net effect of attaching an even number of magnetic flux quanta to every electron. Such composite objects obey the Pauli principle and feel a smaller effective magnetic field (in fact, one which mimics that of an integer quantum Hall effect); to this end, flux attachment provides a unified description for, and explains the similarity between, the experimental results for the integer and fractional cases.

\begin{figure}[H]
\includegraphics[width=3.4in]{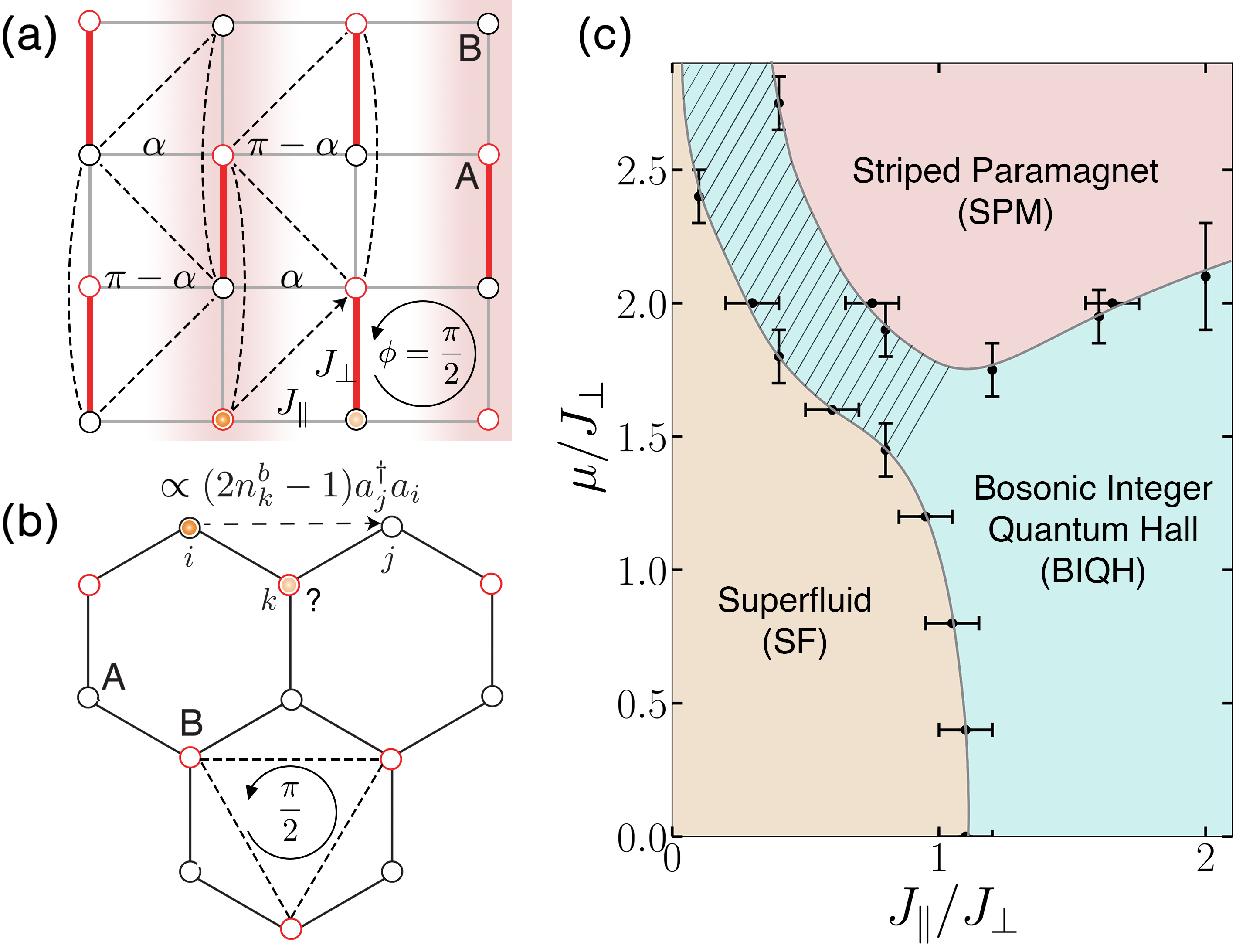} \caption{\label{fig:model} (a) Schematic depiction of our square-lattice, correlated-hopping model for two species of hard-core bosons. 
The background flux of the nearest-neighbor Floquet-driven Hamiltonian is $\phi$ per square plaquette, leading to an effective Hamiltonian with staggered flux $\pi/2 \pm \phi$ per triangular loops shown with dotted lines. The added flux $\alpha$ breaks the time-reversal symmetry of the effective Hamiltonian. The striped chemical potential (red shading), $V=\pm \mu$,  facilitates adiabatic preparation.
(b) Analogous model for a honeycomb lattice, where the background flux is $\pi/2$ per triangular plaquette ($\pi$ per hexagon).
(C) Phase diagram of the square lattice model with $\phi=\pi/2$ at half filling computed via iDMRG on a cylinder of width $L=10$ lattice sites. 
The hashed region appears to be smoothly connected to the BIQH phase. The error bars originate from the resolution of the numerics and the flow of the transition with bond dimension. }
\end{figure}

A considerable amount of recent attention has focused on correlated hopping~\cite{Greschner2014b,di2014quantum,hudomal2020quantum,lienhard2020realization,chu2023photon} of the form,  $H = (2n_k^b-1) a_j^\dag a_i$ [Fig.~\ref{fig:model}(a,b)], in part because it provides a natural framework for implementing flux attachment~\cite{He2015, Fuji2016, Liu2019}.
%
Within this Hamiltonian setting, the $a$-particles naturally see the density of $b$-particles as flux since,  $e^{i\pi n^{b}_k} = (1 - 2n_k^b)$. In addition to enabling a more direct mapping between analytic predictions and microscopic models, such correlated-hopping Hamiltonians have also

\noindent been shown to exhibit (fractional) Chern insulators at anomalously large background fluxes~\cite{He2015}; understanding the interplay between topology and lattice symmetries in this high-flux regime is the subject of active investigation~\cite{regnault2011fractional,murthy2012hamiltonian,zhao2021switching,chen2012majorana,liu2022recent,lin2023complex}. 

Despite seminal advances ~\cite{Liu2019,Greschner2014, Greschner2014b}, owing to the multi-body nature of the interactions, it remains an open challenge to directly implement correlated hopping. 
In this Letter, we propose and analyze a method to realize correlated hopping in a periodically-driven (Floquet) system of spins or hard-core bosonic particles.
Our main results are threefold.
First, we analytically illustrate the emergence of correlated hopping from the periodic modulation~\cite{Dunlap1986,Aidelsburger2011, Struck2012, JimenezGarcia2012, Jotzu2014} of a simple hard-core boson model.
We utilize a perturbative coupled-wire construction to explore the existence of topological phases in the resulting many-body Hamiltonian.
Second, guided by this analysis, we perform large-scale density matrix renormalization group (DMRG) simulations~\cite{tenpy}, which reveal the existence of both a bosonic integer quantum Hall (BIQH) phase on the square lattice~\cite{He2015, Fuji2016, Liu2019,Miao2022}, and a bosonic fractional quantum hall (BFQH) phase on the honeycomb lattice~\cite{Fuji2016,hudomal2019bosonic,leonard2023realization}.
Surprisingly, we also discover a regime where the Hamiltonian is explicitly time-reversal invariant, and yet, hosts a robust BIQH ground state; in this regime, BIQH states with either sign of the Hall conductance are simultaneously stable~\cite{He2015KagomeSPT}. 

Finally, motivated by the possibility of adiabatically preparing the BIQH in cold atomic systems~\cite{barkeshli2015continuous,motruk2017phase}, we explore the surrounding phase diagram  as a function of two natural experimental control parameters: the anisotropy of the hopping strengths, and an overlaid striped chemical potential. 
We provide a specific experimental blueprint for realizing our protocol in a lattice gas of ultracold bosonic atoms~\cite{goldman2014light,dalibard2015introduction,zhai2015degenerate,goldman2016topological,cooper2019topological} \footnote{we emphasize that our protocol can also  be implemented in Rydberg tweezer arrays where synthetic gauge fields arise from either dipolar exchange interactions~\cite{yao2013realizing,cesa2013artificial,lienhard2020realization,nishad2023quantum} or local dressing fields~\cite{celi2014synthetic,wu2022manipulating}.}.
In addition to providing a microscopic route to both realizing and understanding flux attachment, our approach opens the door to a more general framework for defect-particle binding and the simulation of exotic phases and phase transitions~\cite{Senthil2004,Senthil2013,Xu2013, Liu2014}.

\begin{figure}
\includegraphics[width=3.4in]{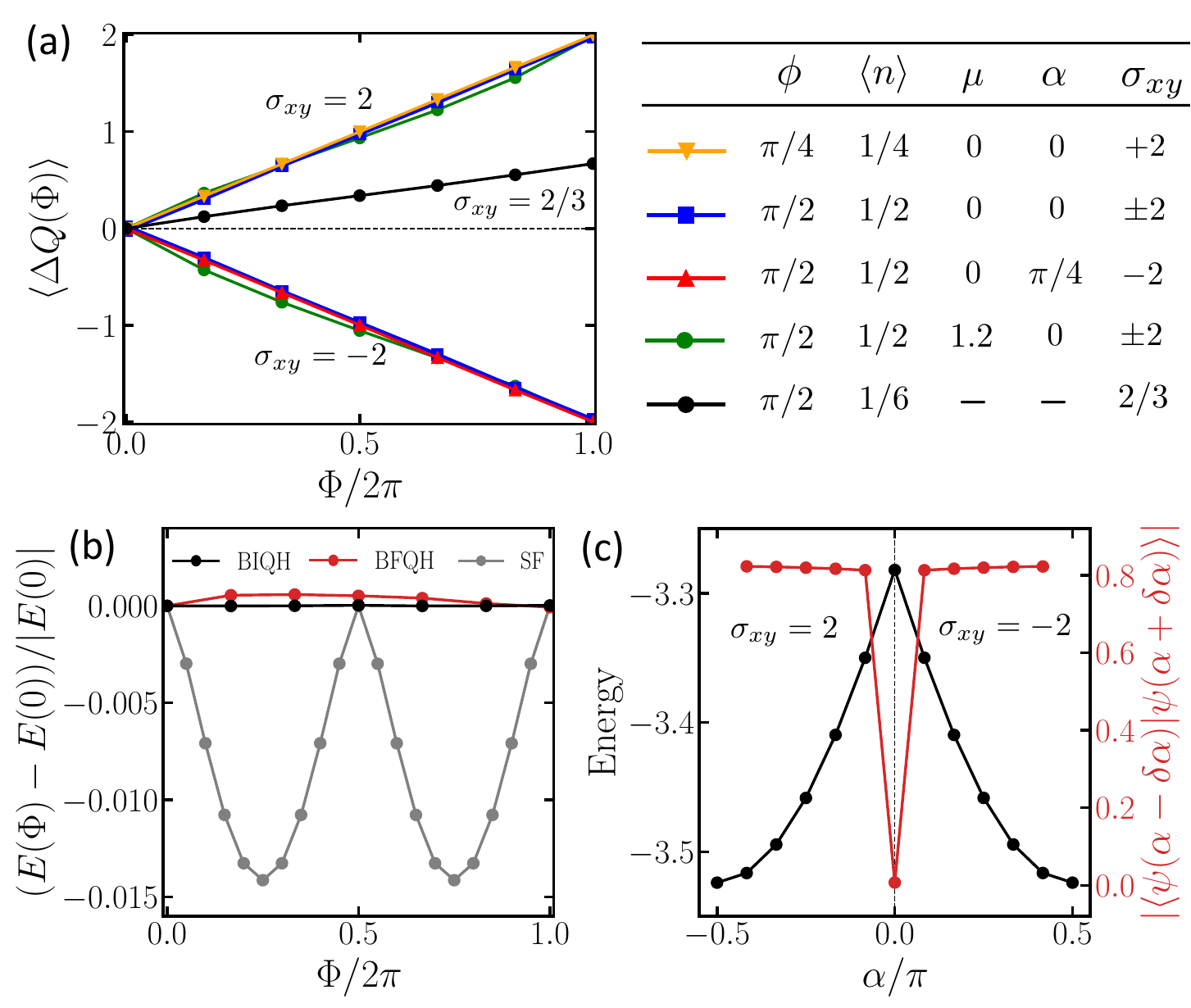}
\caption{\label{fig:numerics} (a) Charge pumping under flux insertion provides numerical evidence for the BIQH state on the square lattice at various parameters, as well as the Halperin-221 fractional quantum Hall state on the isotropic honeycomb lattice (last row of the table). $J_\parallel/J_\perp = 2$ for all the 
BIQH states. The BIQH states were observed on 
cylinder widths $L=6,8,10$ lattice sites, and the BFQH state on a cylinder of width $L=12$ sites. (b) The negligible change in the ground-state energy under flux insertion in the BFQH and half-filled BIQH phases, contrasted with that in the superfluid phase on the square lattice at $J_\parallel/J_\perp = 0.4$. (c) Energy and 
wavefunction overlap as a function of time-reversal symmetry-breaking parameter $\alpha$, showing a first-order phase transition between the two BIQH phases with $\sigma_{xy}=\pm 2$, 
for $\phi=\pi/2, \langle n \rangle =1/2$, $J_\parallel/J_\perp = 2$, and $\delta\alpha=\pi/24$.}
\end{figure}

\emph{Floquet Flux Attachment}---Let us start by demonstrating how a periodically-driven system of free hard-core bosons on a bipartite lattice can generate correlated hopping~[Fig.~\ref{fig:model}(a,b)]. 
%
Consider the Hamiltonian
\begin{equation}\label{eq:NN_time}
H(t)=\sum_{\langle i k \rangle} [J_{ik} \cos(\Omega t + \theta_{ik}) e^{i{\mathcal B}_{i k}}a^\dag_i b_k+\text{H.c.}],
\end{equation}
where $a$ ($b$) is the annihilation operator for a hard-core boson on sublattice $A$ ($B$) of the bipartite lattice, ${\mathcal B}_{ik}$  captures a background flux, $\theta_{ik}$ are bond-dependent constants, and the nearest-neighbor hopping amplitudes are periodically modulated at frequency $\Omega$. Even though the bosons on each sublattice are identical, we distinguish them, as they will be conserved independently in the final effective Hamiltonian.
One can factor out the periodic drive, $H(t)=e^{i\Omega t} H_1+e^{-i\Omega t} H_{-1}$, 
where $H_1= \sum_{\langle i k\rangle} J_{ik} e^{i\theta_{ik}}  [e^{i{\mathcal B}_{ik}}a^\dag_{i} b_k+\text{H.c.}]$ and $H_{-1}= \sum_{\langle i k\rangle} J_{ik} e^{-i\theta_{ik}}  [e^{i{\mathcal B}_{ik}}a^\dag_{i} b_k+ \text{H.c.}]$.
For large driving frequencies, the Hamiltonian can be expanded in powers of $1/\Omega$ using a Floquet-Magnus expansion; the leading-order term is given by \cite{goldman2014periodically}:
\begin{align} \label{eq:Eff_Fl_Ham}
&H_{\rm eff}=\frac{1}{\Omega} [H_1, H_{-1}] \nonumber \\
&=\frac{ 2}{\Omega} \sum_{ij;k} J_{ik}J_{kj}\sin(\theta_{ik}-\theta_{kj}) \left[e^{i ({\mathcal B}_{ij}+\frac{\pi}{2})} (2 n_k^b-1) a^\dag_i a_j+\text{H.c.}\right] \nonumber  \\
& + \frac{ 2}{\Omega} \sum_{kl;i}  J_{ki}J_{il} \sin(\theta_{ki}-\theta_{il}) \left[e^{i ({\mathcal B}_{kl}+\frac{\pi}{2})}(2 n_i^a-1) b^\dag_k b_l + \text{H.c.}\right].
\end{align}
Crucially, $H_{\rm eff}$ exhibits  correlated hopping---as the $a$-bosons hop on the $A$ sublattice, they acquire a phase depending on the occupation of the intervening $B$-lattice site ($n_k^b=b^\dag_k b_k=0, 1$), and vice versa \footnote{This effective static Hamiltonian is prethermal and describes the system for exponentially long times $\sim e^{\Omega/J}$, before drive-induced Floquet heating occurs~\cite{Bukov2016,kuwahara2016floquet,mori2016rigorous,Abanin2017,weidinger2017floquet,machado2019exponentially}.}.

\emph{Bosonic integer quantum Hall}---Consider Eq.~(\ref{eq:Eff_Fl_Ham}) on the square lattice, where the hopping amplitudes along the horizontal and vertical directions are given by $J_\parallel$ and $J_\perp$, respectively [Fig.~\ref{fig:model}(a)]. 
Using a coupled-wire construction, we investigate the existence of interesting quantum Hall states~\cite{Fuji2016, Fuji2019}. 
To facilitate this approach, we choose $\theta_{ik}=\pi/2$ on the thick red bonds depicted in Fig.~\ref{fig:model}(a), while $\theta_{ik}=0$ otherwise. The consequence of this choice for $H_\textrm{eff}$ is that bosons can only hop to their next-nearest neighbors vertically and diagonally, with coupling strengths proportional to $J_\perp^2$ and $ J_\perp J_\parallel$, respectively. Thus, 
when the ratio $J_\parallel/ J_\perp$ vanishes, the bosons cannot hop between vertical chains. A simple Jordan-Wigner transformation reveals that the chains decouple into gapless Luttinger liquids. 
Turning on the diagonal hopping between chains
gaps out the bulk degrees of freedom, but a perturbative analysis reveals that gapless modes can survive at the edge, suggesting a quantum Hall state~\cite{AppendixA, SM}.
In particular, we find that for a system with boson density $\langle n \rangle$ per site and background flux $\phi=\sum_{ij} {\mathcal B}_{ij}=\pi (2p+2q+1)\langle n \rangle$ per plaquette ($p,q \in \mathbb{Z}$), it is possible to realize a Halperin state, where the commutation relations between the gapless modes are described by the  Chern-Simons $K$-matrix: $K = \begin{pmatrix} 2p && 2q+1 \\ 2q+1 && 2p \end{pmatrix}$~\cite{Wen1995, Lu2012}.

The simplest possible such state is the BIQH, a symmetry-protected topological phase 
characterized by $K = \begin{pmatrix} 0 && 1 \\ 1 && 0 \end{pmatrix}$. Guided by the above analysis, we use iDMRG to compute the ground state at filling factor $\nu=2\pi \langle n \rangle/\phi=2$, where $\phi=\pi/4$ and $\langle n \rangle =1/4$, in the hope of observing a BIQH state.  We choose a large value of the ratio $J_\parallel/J_\perp = 2$, such that the system should presumably be deep in this topological phase.
The system is wrapped around an infinitely long cylinder with the vertical ($J_\perp$) and horizontal ($J_\parallel$) hoppings in the wrapping and infinite direction, respectively. We confirm that the choice of wrapping direction has no qualitative effect. 

\begin{figure}
\includegraphics[width=3.4in]{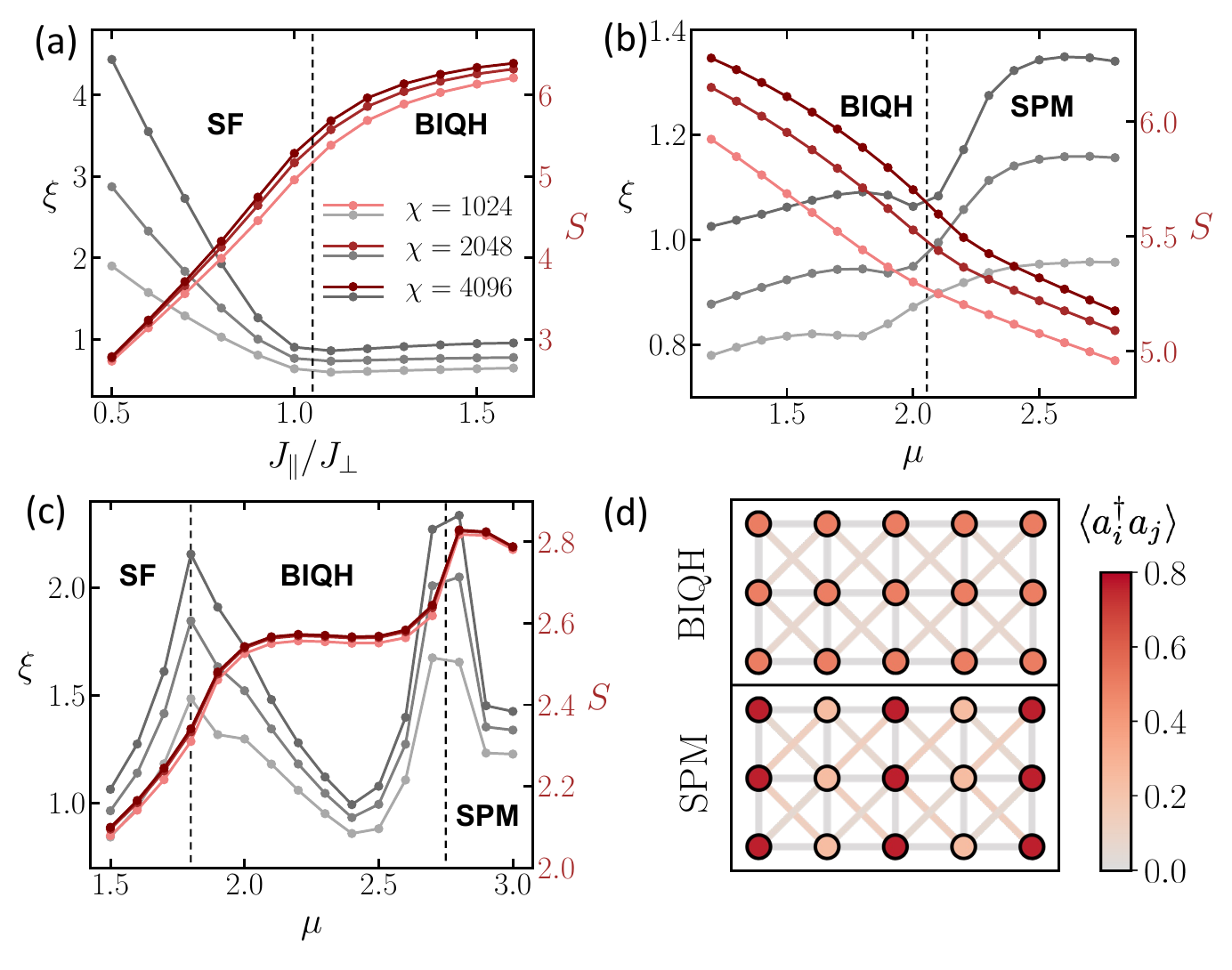}
\caption{\label{fig:numerics2} Phase transitions out of the spontaneous time-reversal symmetry-breaking BIQH phase ($L=10$): correlation length $\xi$ along the cylinder direction and entanglement entropy $S$: (a) at fixed $\mu = 0$, showing a phase transition from a superfluid to BIQH, 
(b) at fixed $J_\parallel/J_\perp = 2.0$, showing a phase transition from BIQH to a striped paramagnetic phase, 
(c) at fixed $J_\parallel/J_\perp = 0.4$, showing the various phase transitions from a superfluid to the striped paramagnetic phase. 
The intervening phase appears to be smoothly connected to the BIQH. (d) Density and nearest-neighbor correlations for the BIQH at $J_\parallel/J_\perp = 2.0, \mu = 0$ and the SPM at $J_\parallel/J_\perp = 2.0, \mu=2.4$. }
\end{figure}

A tell-tale signature of the BIQH phase is its Hall conductance $\sigma_{xy}$, which is always quantized to an even integer. 
Within iDMRG, in order to compute $\sigma_{xy}$~\cite{Laughlin1981}, we thread $2\pi$ flux through the cylinder and measure the resulting charge pumping~\cite{He2014a, Grushin2015}.
As shown in Fig.~\ref{fig:numerics}(a), the charge  pumped increases linearly as a function of the inserted flux, $\Delta Q \equiv  \frac{\Phi}{2\pi} \sigma_{xy}=  \frac{\Phi}{\pi}$, providing evidence that the ground state 
is a BIQH state with a quantized Hall conductance, $\sigma_{xy}=2$. Our numerics suggest that this BIQH state is stable for any $J_\parallel/J_\perp \gtrsim 0.3$.
Moreover, the ground state energy remains nearly constant under flux insertion, indicating  that the system is gapped [Fig.~\ref{fig:numerics}(b)].
The large entanglement entropy and short correlation length provide further evidence that the ground state is indeed a gapped, entangled liquid~[Figs.~\ref{fig:numerics2}(a)]. 
%
%

\emph{Bosonic fractional quantum Hall}---Our coupled-wire analysis suggests that even more exotic states can be realized. For example, choosing $p=1, q=0$ yields the 221-Halperin state~\cite{He2015, Fuji2016}, which is topologically ordered, exhibits a fractional Hall conductance $\sigma_{xy}=2/3$ and supports chiral edge modes, unlike the BIQH state.  
Unfortunately, we were unable to stabilize such a state on the square lattice. However,  a similar coupled-wire analysis on the honeycomb lattice [Fig~\ref{fig:model}(b)] predicts the emergence of a 221-Halperin ground state for the parameters: $J_{ik} = J$, $\langle n \rangle =1/6$,  $\theta_{ik} = 0, 2\pi/3, 4\pi/3$ and a background flux of $\pi/2$ per triangular plaquette ~\cite{He2015, Fuji2016}. 
To investigate this prediction, we again perform iDMRG and measure charge pumping as a function of
flux insertion.
As depicted in Fig.~\ref{fig:numerics}(a), we indeed 
observe the expected $\sigma_{xy}=2/3$.
We note that the ground-state energy exhibits a weak dispersion as a function of flux insertion [Fig.~\ref{fig:numerics}(b)], albeit significantly smaller than the superfluid that we will soon discuss. 

\emph{Spontaneous time-reversal symmetry breaking}---A naive interpretation of our coupled-wire analysis might 
suggest that a BIQH  phase should be stable for \emph{even larger} external flux $\phi=\pi/2$ and filling $\langle n \rangle =1/2$. 
However, at these parameters, 
the coupled-wire analysis is unable to distinguish between the two different BIQH phases with Hall conductance $\sigma_{xy} = \pm 2$. This is because the effective correlated-hopping model [Eq.~\eqref{eq:Eff_Fl_Ham}] is time-reversal invariant for external flux $\phi=\pi/2$---an emergent symmetry which is broken by higher-order terms in the Floquet-Magnus expansion.

Due to this emergent time-reversal symmetry, it is natural to assume that the ground state for the effective model must be time-reversal invariant. 
However, iDMRG instead finds two degenerate ground states with Hall conductance $\sigma_{xy} = \pm 2$, which break time-reversal symmetry  [Fig.~\ref{fig:numerics}(a)]. Interestingly, this model realizes a BIQH state at an unusually large background flux, $\phi = \pi/2$.
Indeed, other models which host the BIQH state, such as the bosonic Harper-Hofstadter model, generally require a smaller flux, which is closer to the continuum limit~\cite{sterdyniak2015bosonic,he2017realizing}.
In contrast, constructing a quantum Hall state by exploiting correlated hopping to directly drive flux attachment allows for the realization of the BIQH state in a more lattice-dominated regime.
This degeneracy can be lifted by applying an additional staggered flux, $\pm \alpha$, per next-nearest neighbor triangular plaquette [Fig.~\ref{fig:model}(a)]. The original, time-reversal invariant point at $\alpha=0$ can then be reinterpreted as the phase-transition point separating the two BIQH phases. If the phase transition is continuous, one expects a time-reversal invariant critical ground state~\cite{He2015}. Instead, we find strong evidence from both the energy and wavefunction overlap [Fig. \ref{fig:numerics}(c)] that the transition is first order, with concurrent spontaneous time-reversal symmetry-breaking at $\alpha=0$.  

\emph{Phase diagram for adiabatic preparation}---In order to explore the possible adiabatic preparation of this spontaneous time-reversal symmetry-breaking BIQH state, we identify two natural tuning parameters, which can drive the system into nearby phases exhibiting lower entanglement. 
In particular, we construct the phase diagram on the square lattice surrounding the BIQH state
as a function of: (i) the hopping anisotropy $J_\parallel/J_\perp$, and (ii) a  striped chemical potential, $V = \pm \mu$ [Fig. \ref{fig:model}(a)]. 

Let us begin by setting $\mu = 0$ and varying the anisotropy. 
As illustrated in Fig.~\ref{fig:numerics2}(a), the system undergoes a phase transition out of the BIQH phase as  $J_\parallel/J_\perp$ is decreased.

For $J_\parallel/J_\perp \lesssim1$, we observe three features indicative of a superfluid phase: (i) sharp decrease of the entanglement entropy, (ii) rapid growth of the correlation length with bond dimension,
and (iii) sharp peaks in the structure factor. 
Moreover, for a superfluid, flux insertion is expected to frustrate the phase coherence and lead to spectral flow of the ground-state energy 
[Fig.~\ref{fig:numerics}(b)]~\footnote{The superfluid energy has periodicity $\pi$ under flux insertion because for odd sublattice cylinder widths ($L/2$), the ground states for periodic and anti-periodic boundary conditions are exactly degenerate.}. Interestingly, our coupled-wire analysis suggests that the BIQH phase should be stable for any non-vanishing $J_\parallel$, suggesting the possibility that the superfluid region in the phase diagram could vanish in the thermodynamic limit;  the observed superfluid is perhaps stabilized by the energy gap present in our finite-width cylinder geometry.

Let us now turn on the chemical potential, $\mu$. 
By doing so, we explicitly
break translation symmetry along the cylinder axis, while leaving it intact along the perpendicular direction. 
At large $\mu$, one expects the system to be in a paramagnetic state  which conforms to this externally-imposed symmetry breaking.
This is precisely what is observed upon increasing $\mu$ at fixed  $J_\parallel/J_\perp = 2$ [Fig. \ref{fig:numerics2}(b)]; the system remains in the BIQH phase for small values of the chemical potential, and then transitions, at $\mu = 2.0 \pm 0.1$ into a striped paramagnet (SPM), a phase easy to access experimentally. 
Intriguingly, it appears that there is no direct transition from the SPM to the SF phase; instead, there is always
an intervening phase [Fig.~\ref{fig:numerics2}(c)]. This phase appears to be smoothly connected to the BIQH phase, although we cannot verify integer charge pumping in the hashed region [Fig.~\ref{fig:model}(c)] due to its proximity to several phase transitions.

\begin{figure}
\includegraphics[width=0.5\textwidth]{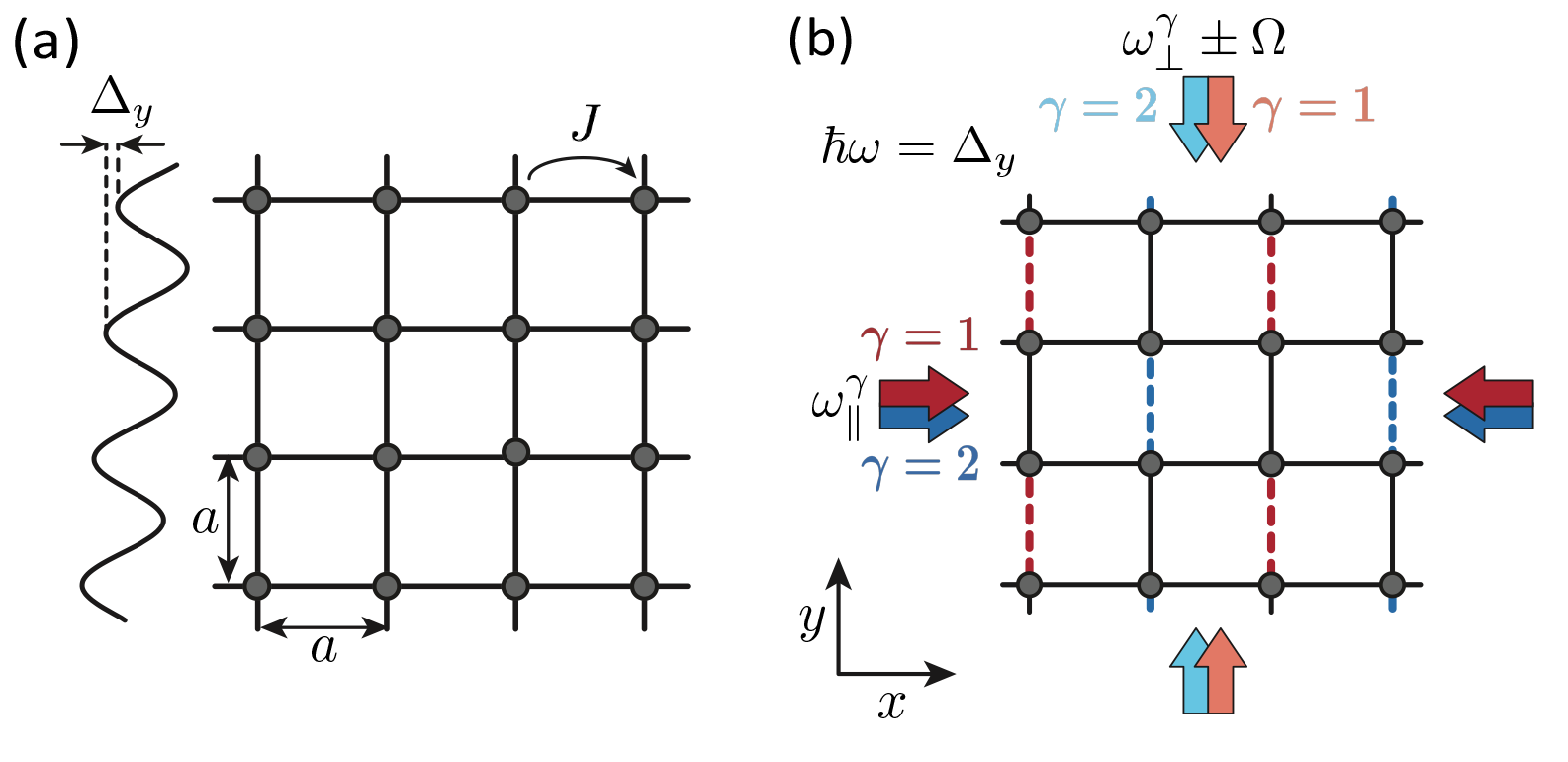} 
\caption{Schematic depiction of the proposed 
setup. (a) 
The vertical hopping is suppressed by an overlaid linear potential $\Delta_y$. (b) The hopping is restored by laser-assisted tunneling. For simplicity, only two of the four laser pairs are shown. 
These allow for Floquet modulation of the correspondingly-colored bonds and control of the background flux.
\label{fig:setup}}
\end{figure}

\emph{Experimental realization}---Motivated by recent advances in the implementation and characterization of topological phases in cold atomic systems~\cite{dalibard2011colloquium,goldman2014light,Jotzu2014,aidelsburger2015measuring,tai2017microscopy,barbiero2019coupling,leonard2023realization, Impertro:2023}, we propose an experimental protocol to directly realize our correlated-hopping model.
In particular, we envision realizing Eq.~(\ref{eq:NN_time}) via laser-assisted tunneling of neutral atoms in a two-dimensional optical lattice~\cite{jaksch2003creation,Aidelsburger2015}.
To be specific, we consider $^{87}$Rb and focus on a square lattice geometry where Harper-Hofstadter models have already been experimentally investigated~\cite{miyake2013realizing,Aidelsburger2013,aidelsburger2015measuring,leonard2023realization}. 
In the presence of a linear potential, $\Delta_y$, along the $y$ axis [Fig.~\ref{fig:setup}(a)], the system is described by the following bosonic tight-binding Hamiltonian:
\begin{align}
    H_0=&-J \sum_{m,n} \left(\hat{a}^{\dagger}_{m+1,n} \hat{a}_{m,n} + \hat{a}^{\dagger}_{m,n+1} \hat{a}_{m,n}+ \text{H.c.} \right)\nonumber \\&+ \sum_{m,n} n \Delta_y \hat{n}_{m,n},
\end{align}
where $m$ ($n$) is the lattice site index along $x$ ($y$). 
For $\Delta_y\gg J$,  tunneling is suppressed along the $y$-axis, while atoms are free to tunnel along the $x$ direction. 

In order to restore resonant tunneling and to realize the background flux $\phi$ [Fig.~\ref{fig:model}(a)], we employ a resonant Floquet modulation scheme. Note that this Floquet modulation is used to realize $H_\textrm{eff}$ from $H_0$ and is distinct from the Floquet flux attachment described by Eq.~(\ref{eq:Eff_Fl_Ham}).
Choosing a set of four independent pairs of laser beams, labeled as $\gamma=\{1,2,3,4\}$ further enables us to realize the bond-dependent phases $\theta_{ik}$. 
Let us illustrate the resonant modulation technique by focusing on the interference generated by a single pair.
The two beams [e.g.~$\gamma=1$ in Fig.~\ref{fig:setup}(b)] are aligned along the $x$ and $y$ axes; both are vertically polarized and retroreflected, with wave vectors  chosen to be $|\mathbf{k}_\parallel|\simeq|\mathbf{k}_\perp|=k=\frac{\pi}{2a}$, where $a$ is the lattice constant. 
This generates a time-dependent interference pattern that can be adjusted relative to the underlying square lattice in order to selectively modulate the colored bonds [Fig.~\ref{fig:setup}(b)]. The 
beams along $y$ have two frequency components, $\omega^{\gamma}_\perp \pm \Omega$, while those along $x$ have a single frequency component, $\omega^{\gamma}_\parallel$; crucially,  the energy difference $\omega^{\gamma}_\perp-\omega^{\gamma}_\parallel=\omega=\Delta_y/\hbar$ is resonant with the potential energy difference between neighboring sites. 

Up to constant terms, the spatially varying, time-dependent potential resulting from each pair of laser beams can be expressed as:
\begin{align}
    V^{\gamma}(t)=&V_0 \cos(\omega t) \cos\left(\Omega t + \theta^{\gamma}_0\right)  \nonumber \\
& \cos\left(m\frac{\pi}{2}+\varphi^{\gamma}_x\right) \cos\left(n\frac{\pi}{2}+\varphi^{\gamma}_y\right),
\end{align}
where $V_0$ is the strength of the potential and $\theta_0^\gamma$ corresponds to the phase of the two sidebands at $\omega_\perp^\gamma \pm \Omega$. 
We specify the following phases for the four distinct pairs of laser-assisted tunneling beams: $\left(\varphi^{\gamma}_x,\varphi^{\gamma}_y\right)=\{(0, \pi/4 ),(\pi/2, -\pi/4), (\pi, 3\pi/4), (\pi/2, \pi/4) \}$.
This allows us to modulate all of the  vertical bonds in the lattice, while maintaining addressability of the phase $\theta_0$. 

In the high-frequency limit, $\omega\gg J/\hbar$,~$\Omega$, and for moderate modulation amplitudes $V_0\lesssim\hbar\omega$, the lowest-order Floquet Hamiltonian of the driven system, 
$H(t)=H_0+\sum_{\gamma}V^{\gamma}(t)$, can be expressed as: 
\begin{align}
        H_\textrm{F}&=-\sum_{m,n,\gamma} [J_{\text{eff}}\cos\left(\Omega t + \theta^{\gamma}_0\right) \hat{a}^{\dagger}_{m,n} \hat{a}_{m,n+1} \cos(m\frac{\pi}{2} +\varphi_x^\gamma)\nonumber \\
        & \hspace{25pt} \sin(n\frac{\pi}{2} + \frac{\pi}{4} +\varphi_y^\gamma) +J \hat{a}^{\dagger}_{m,n} \hat{a}_{m+1,n}+ \text{H.c.} ] ,
        \label{eq:effFloquetExp}
\end{align}
where $J_{\text{eff}}=J\frac{V_0}{\sqrt{2}\Delta_y}$. 
This Hamiltonian realizes a staggered background flux. 
%
To achieve a homogeneous flux, as well as amplitude modulation of the horizontal bonds, we add a staggered potential $\Delta_m = (-1)^m \Delta_x/2$ along the $x$ axis and restore resonant tunneling via an additional pair of running laser beams~\cite{SM,aidelsburger2015measuring}.

To ensure that the leading-order approximations of both the laser Floquet modulation and the Floquet flux attachment are valid, the system must satisfy a hierarchy of energy scales: 
$\omega\gg J/\hbar$,~$\Omega$ and $\Omega\gg J_{\text{eff}}/\hbar$. 
For $^{87}$Rb, the following 
parameters satisfying these criteria can be readily achieved: $\omega=2\pi \times 5\,$kHz, $\Omega=2\pi \times 1\,$kHz and $J_{\text{eff}}/h=100\,$Hz~\cite{Aidelsburger2011,Aidelsburger2015}.

Finally, let us consider the preparation of the BIQH state via a quasi-adiabatic ramp from the striped paramagnet.
Naturally, one expects the maximum length-scale of coherent domains, $l_\mathrm{dec}$, to be limited by the decoherence time, $t_\mathrm{dec}$. 
Although this decoherence time is strongly dependent on the specific platform and driving protocol, current experiments suggest that $t_\mathrm{dec} \sim 100 /J_\perp$ is achievable~\cite{gross2017quantum, jepsen2020spin, ye2022universal, braun2024real}. 
By extracting estimates of typical correlation lengths and energy gaps from our numerical data, combined with the latest quantum Monte Carlo estimates for the critical exponents of the SPM-BIQH transition~\cite{Grover2013,Fuji2016, qin2017duality},  a standard Kibble-Zurek analysis yields $l_\mathrm{dec} \approx 9.5a$~\cite{SM, chandran2012kibble}.
%

\emph{Conclusion---}Our work opens the door to a number of
intriguing directions. 
The correlated-hopping model we consider might be able to stabilize $ppq$-Halperin states beyond the $221$ state at even lower filling, as suggested by our coupled-wire analysis. To realize even more exotic phases, a natural extension of our model would be to relax the hard-core boson constraint, or equivalently to utilize higher-spin Hamiltonians, so that the correlated hopping could drive flux attachment in finer gradations than simply $0$ or $\pi$. More practically, the optimum route for adiabatic preparation of the BIQH or Halperin 221-state, and the timescales required in realistic experimental systems, remain open questions.
\begin{acknowledgments}
\emph{Acknowledgments---}We gratefully acknowledge the insights of and discussions with Philip Crowley, Francisco Machado, Marcus Bintz, Vincent Liu, Dan Stamper-Kurn, David Weld, Michael Lohse, and Immanuel Bloch.
We are particularly indebted to Johannes Hauschild for advice and insights on the TenPy package~\cite{tenpy}.
This work was supported in part by the  Air Force Office of Scientific Research via the MURI program (FA9550-21-1-0069), the ARO via grant no.~W911NF-21-1-0262, by the NSF through the QII-TAQS program (Grant No.~1936100) and by the David and Lucile Packard foundation. M.A.
acknowledges support from the Deutsche Forschungsge-
meinschaft (DFG) via the Research Unit FOR 2414 under
Project No. 277974659 and under Germany’s Excellence Strategy –
EXC-2111 – 390814868. M.A. also acknowledges funding from the European
Research Council (ERC) under the European Union’s Hori-
zon 2020 research and innovation program (grant agreement
No. 803047) and under Horizon Europe programme HORIZON-CL4-
2022-QUANTUM-02-SGA via the project 101113690
(PASQuanS2.1). Y.F. acknowledges support from JSPS KAKENHI Grant No. JP20K14402 and JST CREST Grant No. JPMJCR19T2. Research at Perimeter Institute (Y.C.H) is supported in part by the Government of Canada through the Department of Innovation, Science and Industry Canada and by the Province of Ontario through the Ministry of Colleges and Universities. Work at Innsbruck (P.Z.) supported by the European Union’s Horizon 2020 research and innovation programme under 
grant agreement no. 101113690 (PASQuanS2.1).
\end{acknowledgments}

\emph{Appendix A: Coupled Wire Analysis---} In this Appendix, we provide a field theoretical description of the effective Floquet Hamiltonian Eq.~\eqref{eq:Eff_Fl_Ham} on the square lattice in the limit of $J_\parallel/J_\perp \ll 1$.

For $J_\parallel=0$, vertical chains are decoupled and the corresponding Floquet Hamiltonian for each chain is given by
\begin{align}
H_\textrm{eff} = t_\perp \sum_\ell \big[ &i a^\dagger_\ell a_{\ell+1} (2n^b_\ell-1) -i b^\dagger_\ell b_{\ell+1} (2n^a_{\ell+1}-1)  \nonumber
\\
&+\textrm{H.c.} \big], 
\end{align}
where $t_\perp = 2J_\perp^2/\Omega$.
This Hamiltonian can be mapped onto two decoupled hard-core boson chains with the standard hopping via the Jordan-Wigner transformation \cite{Fuji2016}, 
\begin{align} \label{eq:JWBosons}
a_\ell = (-i)^\ell \ta_\ell \tK^b_\ell, \ \ \ 
b_\ell = (+i)^\ell \tb_\ell \tK^a_\ell,
\end{align}
with the string operators,
\begin{align}
\tK^a_\ell = \cos \left( \pi \sum_{\ell' \leq \ell} \tn^a_{\ell'} \right), \ \ \ 
\tK^b_\ell = \cos \left( \pi \sum_{\ell' \geq \ell} \tn^b_{\ell'} \right).
\end{align}
Note $\tn^{a(b)}_\ell = n^{a(b)}_\ell$. 
We then find 
\begin{align}
H_\textrm{eff} = -t_\perp \sum_\ell \left( \ta^\dagger_\ell \ta_{\ell+1} +\tb^\dagger_\ell \tb_{\ell+1} +\textrm{H.c.} \right).
\end{align}

\begin{figure}
\includegraphics[width=0.4\textwidth]{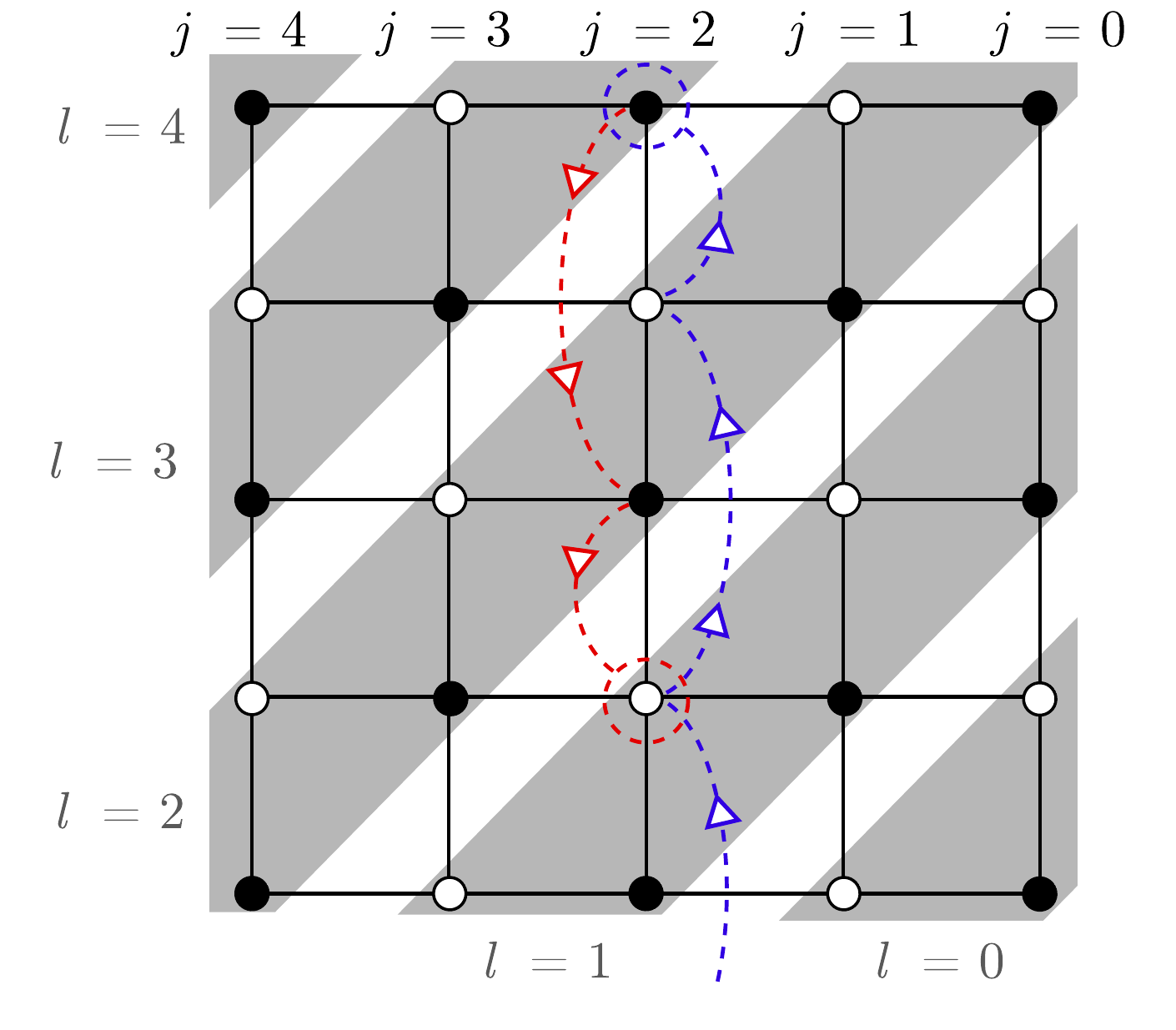}
\caption{\label{fig:coupledwires}
The coordinate system used for the square lattice, labeling vertical chains with $j$ and each second diagonal with $l$.
The Jordan-Wigner string for $a_{1}$ ($b_{3}$), enclosed by the blue (red) circle, is represented by the blue (red) dashed line. }
\end{figure}

Now we can apply the standard bosonization technique~\cite{Giamarchi}. 
The low-energy effective Hamiltonian is given by the free boson theory,
\begin{align}
H_\textrm{eff} \sim \int dx \sum_{s=a,b} \frac{v^s}{2\pi}  \left[ (\partial_x \varphi^s)^2 +(\partial_x \theta^s)^2 \right],
\end{align}
where $v^s = 2t_\perp \sin(\pi \langle n^s \rangle)$, $x=\ell d_0$ with $d_0$ being the lattice spacing, and the bosonic fields satisfy the commutation relations,
\begin{align}
[\theta^s(x), \varphi^{s'}(x')] = i\pi \delta_{ss'} \Theta(x-x'), 
\end{align}
with $\Theta(x)$ being the Heaviside step function. 
The lattice boson operators are expressed in terms of the bosonic fields $\tilde{s}_\ell= \ta_\ell, \tb_\ell$ as
\begin{align}
\label{eq:BosonOpA}
\tilde{s}_\ell &\sim e^{-i\varphi^s(x)} \sum_{m \geq 0} C^s_{2m} \cos \left( 2m \theta^s(x) +2\pi m \langle n^s \rangle \ell \right), \\
\tn^s_\ell &\sim \langle n^s \rangle +\frac{d_0}{\pi} \partial_x \theta^s(x) +\cdots.
\end{align}
The string operators may also be expressed as 
\begin{align}
\tK^s_\ell \sim \sum_{m \geq 0} D^s_{2m+1} \cos \big[ &(2m+1) (\theta^s(x)-\delta_{sb} N^b)\nonumber \\&+\pi (2m+1) \langle n^s \rangle\ell \big], \label{eq:StringOpB}
\end{align}
Here $C^s_{2m}$ and $D^s_{2m+1}$ are numerical constants and $N^s$ are zero-mode operators formally defined by 
\begin{align}
N^s = \int^\infty_{-\infty} dx \frac{1}{\pi} \partial_x \theta^s(x).
\end{align}
Let us consider the single particle correlation functions $\langle a^\dagger_\ell a_{\ell'} \rangle$ and $\langle b^\dagger_\ell b_{\ell'} \rangle$.
In our free boson theory, the vertex operators $e^{i m \varphi^s(x)}$ and $e^{i m \theta^s(x)}$ both have scaling dimension $m^2/4$. 
Using Eqs.~\eqref{eq:JWBosons} and \eqref{eq:BosonOpA}-\eqref{eq:StringOpB} and keeping only most slowly decaying parts, we find 
\begin{align}
\langle s^\dagger_\ell s_{\ell'} \rangle \sim \frac{C^s_0 D^{\bar{s}}_1}{4} \bigg( & \frac{e^{i[(-1)^{\delta_{sb}}\frac{\pi}{2} +\pi \langle n^{\bar{s}} \rangle](\ell-\ell')}}{|\ell-\ell'|} \nonumber
\\& +\frac{e^{i[(-1)^{\delta_{sb}}\frac{\pi}{2} -\pi \langle n^{\bar{s}} \rangle](\ell-\ell')}}{|\ell-\ell'|} \bigg) +\cdots,
\end{align}
where $s=a,b$ and $\bar{s} = b,a$. The momentum distribution functions are obtained by their Fourier transforms: $\langle n^s(k) \rangle = \langle s^\dagger(k) s(k) \rangle$. 
They diverge logarithmically as $k$ approaches $\frac{\pi}{2} \pm \pi \langle n^b \rangle$ for the $A$ sublattice and $-\frac{\pi}{2} \pm \pi \langle n^a \rangle$ for the $B$ sublattice, 
\begin{align}
\langle n^s(k) \rangle \sim \log \frac{1}{|k-(-1)^{\delta_{sb}}\frac{\pi}{2} \mp \pi \langle n^{\bar{s}} \rangle|} . 
\end{align}
The positions of the singularities depend on the boson density of different species.
This behavior is contrasted from that for the standard hard-core boson chain, $H = -t \sum_\ell (a^\dagger_\ell a_{\ell+1} +\textrm{H.c.})$, for which the momentum distribution function exhibits a power-law divergence $\langle n^a(k) \rangle \sim |k|^{-1/2}$ as $k \to 0$ \cite{Giamarchi}. 

We are now in a position to study the low-energy physics of the correlated hopping chains weakly coupled by $J_\parallel$ in the spirit of coupled-wire construction \cite{Kane2002,Teo2014,Lu2012}. 
We start from the decoupled chain Hamiltonian, 
\begin{align}
H^0_\textrm{eff} = t_\perp \sum_{j,\ell} \big[ &ia^\dagger_{j,\ell} a_{j,\ell+1} (2n^b_{j,\ell} -1)\nonumber \\ &-i b^\dagger_{j,\ell} b_{j,\ell+1} (2n^a_{j,\ell+1} -1) +\textrm{H.c.} \big], 
\end{align}
where the site index $(j,\ell)$ is assigned in the way depicted in Fig.~\ref{fig:coupledwires}. 
With this assignment and choosing a Landau gauge to implement the flux $\phi$ for each square plaquette, the effective Floquet Hamiltonian involving $J_\parallel$ couplings is given by 
\begin{align}
H^1_\textrm{eff} =& t_\parallel \sum_{j,\ell} \big[ -i e^{-i\phi(2\ell)} a^\dagger_{j,\ell} a_{j+1,\ell} (2n^b_{j+1,\ell}-1) \nonumber  \\ &+i e^{-i \phi (2\ell+1)} a^\dagger_{j,\ell} a_{j+1,\ell+1} (2n^b_{j,\ell}-1) \nonumber  \\ & +i e^{-i\phi(2\ell)} b^\dagger_{j,\ell} b_{j+1,\ell} (2n^a_{j,\ell}-1)  \nonumber  \\ & -i e^{-i\phi(2\ell+1)} b^\dagger_{j,\ell} b_{j+1,\ell+1} (2n^a_{j+1,\ell+1}-1) +\textrm{H.c.} \big], \nonumber
\end{align}
where $t_\parallel = 2J_\perp J_\parallel/\Omega$. 
After applying the Jordan-Wigner transformation \eqref{eq:JWBosons} for each chain, and treating 
$H^1_\textrm{eff}$ as a perturbation for $|J_\parallel| \ll |J_\perp|$, we can apply the above bosonization procedure to obtain the low-energy effective Hamiltonian as similarly done in Ref.~\cite{Fuji2016}. Keeping only terms relevant for quantum Hall states, we find
\begin{align}
\label{eq:ChainHam}
H^0_\textrm{eff} &\sim \int dx \sum_j \sum_{s=a,b} \frac{v^s}{2\pi} \left[ (\partial_x \varphi^s_j)^2 +(\partial_x \theta^s_j)^2 \right], \nonumber \\
H^1_\textrm{eff} &\sim -\sum_{p \in \mathbb{Z}} \sum_{q \in \mathbb{Z}} \sum_j \int dx \bigg[ g^a_{(p,q)} e^{i\pi (2q+1) \langle n^b \rangle} (-i+e^{-i\Gamma^a_{(p,q)}}) \nonumber \\&  \cdot e^{-2i \Gamma^a_{(p,q)} x/d_0} e^{i \chi^a_{(p,q),j}(x) -i\tchi^a_{(p,q),j+1}(x)}  \nonumber \\
& +g^b_{(p,q)} e^{-i\pi(2q+1) \langle n^a \rangle} (i+e^{-i\Gamma^b_{(p,q)}}) \nonumber  \\& \cdot e^{-2i  \Gamma^b_{(p,q)} x/d_0} e^{i\chi^b_{(p,q),j}(x)-i\tchi^b_{(p,q),j+1}(x)} +\textrm{H.c.} \bigg] +\cdots, 
\end{align}
where the bosonic fields satisfy the commutation relations $[\theta^s_j(x), \varphi^{s'}_{j'}(x')] = i\pi \delta_{ss'} \delta_{jj'} \Theta(x-x')$, and we have defined
\begin{align}
g^s_{(p,q)} &= \frac{t_\parallel (C^s_{|2p|} D^{\bar{s}}_{|2q+1|})^2}{4}, \nonumber\\ 
\Gamma^s_{(p,q)} &= \phi -\pi(2p) \langle n^s \rangle -\pi (2q+1) \langle n^{\bar{s}} \rangle, \nonumber\\ 
\begin{split}
\chi^s_{(p,q),j}(x) &= \varphi^s_j(x) +2p \theta^s_j(x) +(2q+1) (\theta^{\bar{s}}_j(x) -\delta_{sa}\pi N^b_j), \\
\tchi^s_{(p,q),j}(x) &= \varphi^s_j(x) -2p \theta^s_j(x) -(2q+1) (\theta^{\bar{s}}_j(x) -\delta_{sa}\pi N^b_j). 
\end{split} \nonumber
\end{align}
If the couplings $g^a_{(p,q)}$ and $g^b_{(p,q)}$ both flow to the strong-coupling fixed point, it is possible to find a quantum Hall state described by the $2 \times 2$ $K$ matrix \cite{Wen1995}, 
\begin{align}
K = \begin{pmatrix} 2p & 2q+1 \\ 2q+1 & 2p \end{pmatrix}. 
\end{align}
This can be understood by the fact that the bosonic fields $\tchi^s_{(p,q),j}(x)$ and $\chi^s_{(p,q),j}(x)$ are left unpaired and remain gapless at the outermost wires $j=1$ and $N_w$, respectively, while they are gapped elsewhere in the bulk $1<j<N_w$.
These bosonic fields satisfy the commutation relations,
\begin{align}
[\partial_x \chi^s_{(p,q),j}(x), \chi^{s'}_{(p,q),j'}(x')] &= -[\partial_x \tchi^s_{(p,q),j}(x), \tchi^{s'}_{(p,q),j'}(x')] \nonumber\\&= 2i\pi K_{ss'} \delta_{jj'} \delta(x-x'), \nonumber 
\end{align}
as expected from the Chern-Simons theory \cite{Wen1995}. 
To obtain the quantum Hall states, several remarks are in order. 
For given $p$ and $q$, the flux $\phi$ and the boson densities $\langle n^s \rangle$ must satisfy the commensurability condition $\Gamma^s_{(p,q)} \equiv 0$ for the corresponding interactions to slowly vary in $x$. 
This requires
\begin{align}
\begin{split}
\phi &= \pi (2p) \langle n^s \rangle +\pi (2q+1) \langle n^{\bar{s}} \rangle \mod \pi. 
\end{split}
\end{align}
 We also note that in the perturbatively accessible regime $|J_\parallel| \ll |J_\perp|$, only the coupling constants with $(p,q)=(0,0)$ or $(0,-1)$ become relevant in the renormalization group sense. 
This corresponds to the BIQH state, the focus of the main text. 
Indeed, the choice of parameters $\langle n^a \rangle = \langle n^b \rangle \equiv \langle n \rangle = 1/2$ and $\phi= \pi /2$ used for the numerical study meets the above commensurability conditions. However, at the small system sizes accessible to numerics, we have of course not yet reached the strong-coupling fixed point. This explains why at perturbatively small $J_\parallel/J_\perp$ we observe a superfluid phase, but at large enough $J_\parallel/J_\perp$ there is a phase transition into a BIQH state.

\bibliography{refs.bib}

\onecolumngrid
\clearpage

\widetext

\setcounter{equation}{0}
\setcounter{figure}{0}
\setcounter{table}{0}
\setcounter{page}{1}
\makeatletter
\renewcommand{\theequation}{S\arabic{equation}}
\renewcommand{\thefigure}{S\arabic{figure}}
\renewcommand{\bibnumfmt}[1]{[S#1]}
\renewcommand{\citenumfont}[1]{S#1}
\newcommand{\sgn}{\operatorname{sgn}}
\newcommand{\eff}{\mathrm{eff}}
\newcommand{\local}{{\mathrm{local}}}
\newcommand{\static}{{\mathrm{static}}}
\newcommand{\drive}{{\mathrm{drive}}}
\newcommand{\expct}[1]{\langle #1 \rangle}
\def\X{\sigma^x}
\def\Y{\sigma^y}
\def\Z{\sigma^z}
\def \ab{\alpha^\prime}
\def \abb{\alpha^{\prime\prime}}
\def \bb{\beta^\prime}
\def \bbb{\beta^{\prime\prime}}
\def \nop{-1^{\prime}}

\begin{center}
\textbf{\large Supplementary Information:\\\vspace{1mm} Floquet Flux Attachment in Cold Atomic Systems}

\vspace{10pt}
\thispagestyle{plain}

\mbox{Helia Kamal,\textsuperscript{1} Jack Kemp,\textsuperscript{1} Yin-Chen He,\textsuperscript{2} Yohei Fuji,\textsuperscript{3} Monika Aidelsburger,\textsuperscript{4,5} Peter Zoller,\textsuperscript{6,7} and Norman Y. Yao\textsuperscript{1}}

\vspace{4pt}
\textsuperscript{1}\textit{\small Department of Physics, Harvard University, Cambridge, MA 02138, USA}

\textsuperscript{2}\textit{Perimeter Institute for Theoretical Physics, Waterloo, ON N2L 2Y5, Canada}

\textsuperscript{3}\textit{Department of Applied Physics, University of Tokyo, Tokyo 113-8656, Japan}
\textsuperscript{4}\textit{Faculty of Physics, Ludwig-Maximilians-Universit\"{a}t M\"{u}nchen, Schellingstr. 4, D-80799 Munich, Germany}
\textsuperscript{5}\textit{Munich Center for Quantum Science and Technology (MCQST), Schellingstr. 4, D-80799 Munich, Germany}

\textsuperscript{6}\textit{Institute for Theoretical Physics, University of Innsbruck, Innsbruck, 6020, Austria}

\mbox{\textsuperscript{7}\textit{Institute for Quantum Optics and Quantum Information of the Austrian Academy of Sciences, Innsbruck, 6020, Austria}}

\end{center}
\section{Details of the Floquet-Magnus expansion}
Here we will provide the details for the Floquet-Magnus expansion used to derive the effective Hamiltonian Eq. 2 in the main text. Firstly, note that
\begin{align}
& [a^\dag_{k+\vec e_\alpha} b_k+b^\dag_k a_{l+\vec e_\beta}, a^\dag_{l+\vec e_\beta} b_l+b^\dag_l a_{l+\vec e_\beta}]  \nonumber
\\ =& [a^\dag_{k+\vec e_\alpha} b_k, b^\dag_l a_{l+\vec e_\beta}]+[b^\dag_k a_{l+\vec e_\beta}, a^\dag_{l+\vec e_\beta} b_l]  \nonumber
\\ = &\delta_{k,l}  [b_k, b^\dag_l] a^\dag_{k+\vec e_\alpha} a_{l+\vec e_\beta}+\delta_{k+\vec e_\alpha, l+\vec e_\beta} [a^\dag_{k+\vec e_\alpha}, a_{l+\vec e_\beta}] b^\dag_l b_k
+\delta_{k,l}  [b_k^\dag, b_l] a^\dag_{l+\vec e_\beta} a_{k+\vec e_\alpha}+\delta_{k+\vec e_\alpha, l+\vec e_\beta} [a_{k+\vec e_\alpha}, a^\dag_{l+\vec e_\beta}] b^\dag_k b_l.
\end{align}
Then we have
\begin{align}
&\sum_k \sum_{\alpha \neq \beta } e^{i\theta_\alpha-i\theta_\beta} [b^\dag_k, b_k] (e^{i A^0_{k+\vec e_\beta, k}+i A^0_{k, k+\vec e_\alpha}}a^\dag_{k+\vec e_\beta} a_{k+\vec e_\alpha}-e^{-i A^0_{k+\vec e_\beta, k}-i A^0_{k, k+\vec e_\alpha}}a^\dag_{k+\vec e_\alpha} a_{k+\vec e_\beta}) \nonumber \\
=&\sqrt 3 \sum_k (2n_k^b-1) \sum_{\alpha=1,2,3} [e^{i A^0_{k+\vec e_{\alpha+1}, k+\vec e_{\alpha}}+i \pi/2} a^\dag_{k+\vec e_{\alpha+1}} a_{k+\vec e_\alpha}+\text{H.c.}]. \label{eq:S2}
\end{align}
Setting $k+\vec e_\alpha=l+\vec e_\beta=i$, 
\begin{align}
& \sum_i \sum_{\alpha \neq \beta} e^{i\theta_\alpha-i\theta_\beta } [a^\dag_i, a_i] (e^{-i A^0_{i-\vec e_\alpha, i}-iA^0_{i, i-\vec e_\beta}}b^\dag_{i-\vec e_\beta} b_{i-\vec e_\alpha}- e^{i A^0_{i-\vec e_\alpha, i}+iA^0_{i, i-\vec e_\beta}}b^\dag_{i-\vec e_\alpha} b_{i-\vec e_\beta})    \nonumber \\
=& \sqrt 3 \sum_i (2n_i^a-1) \sum_{\alpha=1,2,3} [e^{i A^0_{i-\vec e_{\alpha+1}, i-\vec e_{\alpha}}+i \pi/2} b^\dag_{i-\vec e_{\alpha+1}} b_{i-\vec e_\alpha}+\text{H.c.}] \label{eq:S3}
\end{align}
Substituting equations \eqref{eq:S2} and \eqref{eq:S3} back in to the first line of Eq. 2 yields the correct expression for the effective Hamiltonian at leading order in the Floquet-Magnus expansion.

\section{Mutual Flux Attachment in our Coupled-wire system}

In this appendix, we show that the mutual flux attachment proposed in Ref.~\cite{sSenthil2013} is nicely furnished in our coupled-wire system for the BIQH state (provided in Appendix A of the main text) without any uncontrollable approximation. 
The following argument is strongly inspired from a recent study of duality web in Ref.~\cite{sMross2017}. 
We now define the ``mutual composite boson'' fields by a nonlocal transformation~\cite{sFuji2019}, 
\begin{align}
\Phi^a_j(x) &= \varphi^a_j(x) +\sum_{j' \neq j} \textrm{sgn}(j'-j) (\theta^b_{j'} (x)-\pi N^b_{j'}), \\
\Phi^b_j(x) &= \varphi^b_j(x) +\sum_{j' \neq j} \textrm{sgn}(j'-j) \theta^a_{j'}(x), \\
\Theta^s_j(x) &= \theta^s_j(x). 
\end{align}
They satisfy the commutation relations, 
\begin{align} \label{eq:CompositeBosComm}
\begin{split}
[\Theta^a_j(x), \Phi^a_{j'}(x')] &= i\pi \delta_{jj'} \Theta(x-x'), \\
[\Theta^b_j(x), \Phi^b_{j'}(x')] &= i\pi \delta_{jj'} [\Theta(x-x')-1]
\end{split}
\end{align}
while the other commutators vanish. 
We then consider the Euclidean action corresponding to the decoupled chain Hamiltonian Eq. 19, 
\begin{align}
S_0 = \int d\tau dx \sum_j \sum_{s=a,b} \left[ \frac{i}{\pi} \partial_x \theta^s_j \partial_\tau \varphi^s_j +\frac{v^s}{2\pi} \left\{ (\partial_x \varphi^s_j)^2 +(\partial_x \theta^s_j)^2 \right\} \right]. 
\end{align}
In terms of the mutual composite boson fields, the action is written as 
\begin{align}
S_0 = \int d\tau dx \sum_j \sum_{s=a,b} \biggl[ \frac{i}{\pi} \partial_x \Theta^s_j \partial_\tau \Phi^s_j +\frac{v^s}{2\pi} \Bigl\{ \Bigl( \partial_x \Phi^s_j -\sum_{j' \neq j} \textrm{sgn}(j'-j) \partial_x \Theta^{s'}_j \Bigr)^2 +(\partial_x \Theta^s_j)^2 \Bigr\} \biggr], 
\end{align}
where $s'=b(a)$ for $s=a(b)$. 
To formally resolve the nonlocality of this action, we define auxiliary fields $a^1_{1,j}$ and $a^2_{1,j}$ by 
\begin{align}
a^1_{1,j} = \sum_{j' \neq j} \textrm{sgn}(j'-j) \partial_x \Theta^b_{j'}, \ \ \ 
a^2_{1,j} = \sum_{j' \neq j} \textrm{sgn}(j'-j) \partial_x \Theta^a_{j'}, 
\end{align}
and implement these constraints by Lagrange multipliers $a^1_{0,j+\frac{1}{2}}$ and $a^2_{0,j+\frac{1}{2}}$ as
\begin{align}
S_\textrm{LM} &= \int d\tau dx \sum_j \frac{i}{2\pi} \biggl[ (a^1_{0,j+\frac{1}{2}} -a^1_{0,j-\frac{1}{2}}) \Bigl\{ a^2_{1,j} -\sum_{j' \neq j} \textrm{sgn}(j'-j) \partial_x \Theta^a_{j'} \Bigr\} \nonumber \\
&+(a^2_{0,j+\frac{1}{2}} -a^2_{0,j-\frac{1}{2}}) \Bigl\{ a^1_{1,j} -\sum_{j' \neq j} \textrm{sgn}(j'-j) \partial_x \Theta^b_{j'} \Bigr\} \biggr]. 
\end{align}
After some algebra, the action $S'_0 = S_0 +S_\textrm{LM}$ is expressed as
\begin{align}
S'_0 &= \int d\tau dx \sum_j \Biggl[ \frac{i}{\pi} \partial_x \Theta^a_j \biggl( \partial_\tau \Phi^a_j -\frac{a^1_{0,j+\frac{1}{2}} +a^1_{0,j-\frac{1}{2}}}{2} \biggr) + \frac{i}{\pi} \partial_x \Theta^b_j \biggl( \partial_\tau \Phi^b_j -\frac{a^2_{0,j+\frac{1}{2}} +a^2_{0,j-\frac{1}{2}}}{2} \biggr) \nonumber \\
&+\frac{v^a}{2\pi} \Bigl\{ (\partial_x \Phi^a_j -a^1_{1,j})^2 +(\partial_x \Theta^a_j)^2 \Bigr\} +\frac{v^b}{2\pi} \Bigl\{ (\partial_x \Phi^b_j -a^2_{1,j})^2 +(\partial_x \Theta^b_j)^2 \Bigr\} \nonumber \\
&+\frac{i}{4\pi} \Bigl\{ a^1_{1,j} (a^2_{0,j+\frac{1}{2}} -a^2_{0,j-\frac{1}{2}}) -a^2_{0,j+\frac{1}{2}} (a^1_{1,j+1} -a^1_{1,j}) \Bigr\} +\frac{i}{4\pi} \Bigl\{ a^2_{1,j} (a^1_{0,j+\frac{1}{2}} -a^1_{0,j-\frac{1}{2}}) -a^1_{0,j+\frac{1}{2}} (a^2_{1,j+1} -a^2_{1,j}) \Bigr\} \Biggr]. 
\end{align}
This action can be seen as a discrete analog of two-component bosonic fields minimally coupled with the mutual Chern-Simons term $\frac{i}{4\pi} \epsilon^{\mu \nu \lambda} (a^1_\mu \partial_\nu a^2_\lambda +a^2_\mu \partial_\nu a^1_\lambda)$ under the gauge choice $a^1_2 = a^2_2 =0$.
We also consider interchain interactions corresponding to the BIQH state with $(p,q)=(0,0)$, 
\begin{align}
H_1 &= -\int dx \sum_j \left[ g^a \cos \left( \varphi^a_j +\theta^b_j -\pi N^b_j -\varphi^a_{j+1} +\theta^b_{j+1} -\pi N^b_{j+1} \right) +g^b \cos \left( \varphi^b_j +\theta^a_j -\varphi^b_{j+1} +\theta^a_{j+1} \right) \right] \nonumber \\
&= -\int dx \sum_j \left[ g^a \cos \left( \Phi^a_j -\Phi^a_{j+1} \right) +g^b \cos \left( \Phi^b_j -\Phi^b_{j+1} \right)\right].
\end{align}
The interactions maintain a local form in terms of the mutual composite boson fields. 
According to Eq.~\eqref{eq:CompositeBosComm}, the operators $e^{i\Phi^s_j(x)}$ can be viewed as bosonic particle operators and thus the interactions can induce a condensation. 
This precisely reproduces the argument of Ref.~\cite{sSenthil2013} which states that the BIQH state is obtained by the condensation of composite bosons with the mutual $2\pi$ flux attachment.

\section{Numerical Methods}
The numerical simulations are carried out using DMRG algorithm of TenPy library~\cite{Stenpy}. The 2D lattice is placed on an infinitely long cylinder of width $L$ sites, with periodic boundary conditions in both direction. For the honeycomb lattice, "Cstyle" ordering of the MPS is used.

Imposing the symmetries of the Hamiltonian by conserving charges in tensor network methods leads to large speedups and reduction in memory usage. However, enforcing too many restrictions could, in turn, result in the algorithm getting stuck in local minima. In our simulations, we find that explicitly imposing the $U(1) \times U(1)$ symmetry, i.e. particle number conservation on each sublattice, would sometimes lead to failure of DMRG in finding the correct ground state. Thus, we conserve the total particle number only. However, we carefully confirmed that in all cases the final ground state obtained does not break the full $U(1) \times U(1)$ symmetry explicitly.

To implement the background flux on the honeycomb lattice, we utilize the same gauge choice as Ref.~\cite{sHe2015} for the next-nearest-neighbor couplings. On the square lattice, we use a Landau gauge for the choice of nearest-neighbor $B_{ik}$, and set $B_{ij} = B_{ik} + B_{kj}$ (as defined in the derivation of the effective Hamiltonian Eq. 2 in the main text) for the the next-nearest-neighbor correlated hoppings $(2n^b_k - 1)a^\dagger_ia_j$.

\section{Effects of system size}

\begin{figure}
\centering
\includegraphics[width=0.6\textwidth]{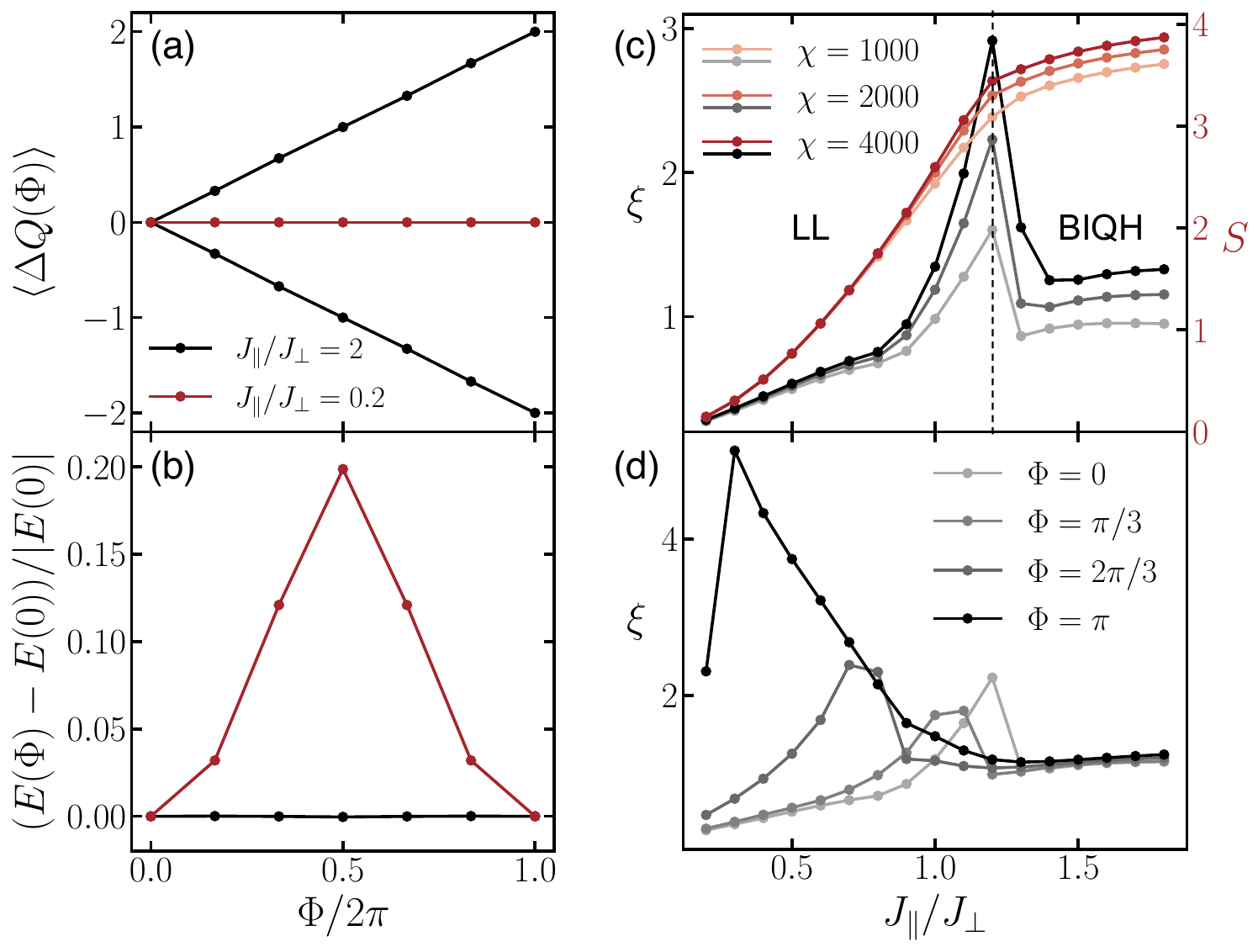}
\caption{\label{fig:BIQH_LL}
Numerical study of the correlated hopping model on the square lattice with half filling and background flux $\phi=\pi/2$, simulated on a cylinder of width $L=8$ sites. (a) Charge pumping under flux threading. (b) Energy response under flux threading. (c) Correlation length and entanglement entropy, showing a phase transition from a Luttinger liquid to BIQH with increasing $J_\parallel/J_\perp$. (d) Shifting of the critical point to smaller values of $J_\parallel/J_\perp$ as the threaded flux increases from $\Phi=0$ to $\pi$.}
\end{figure}

\begin{figure}
\centering
\includegraphics[width=0.9\textwidth]{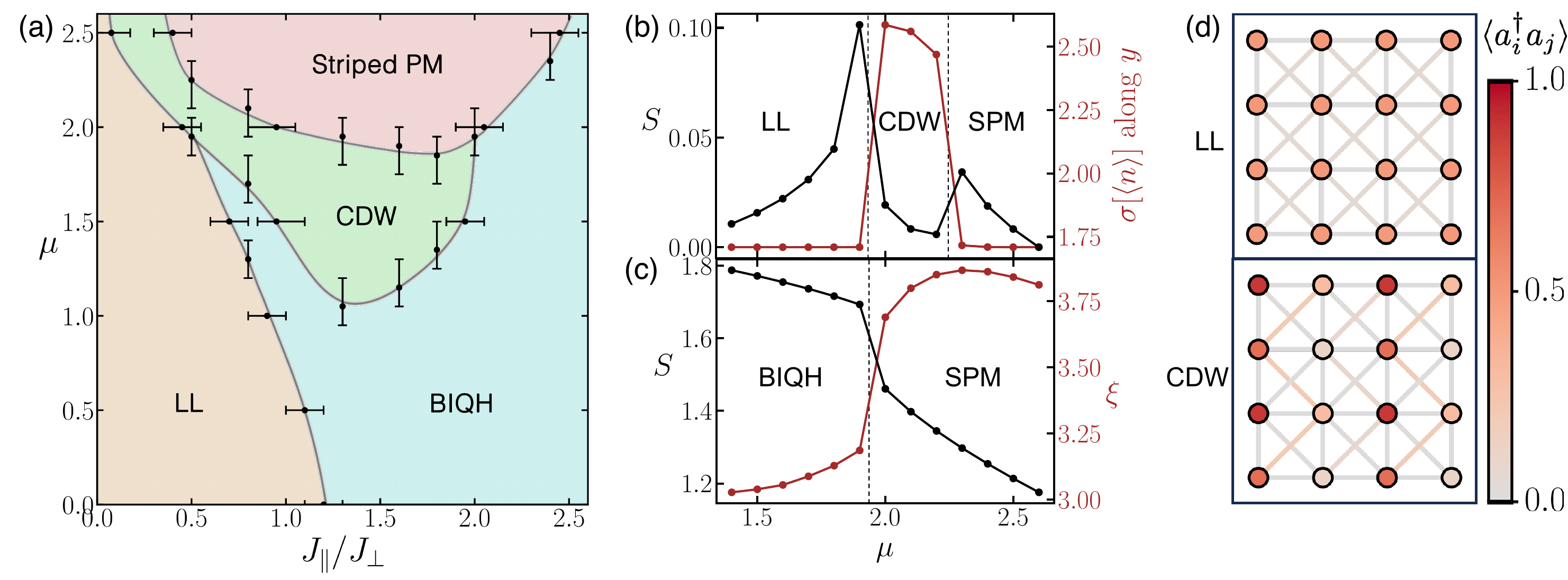}
\caption{\label{fig:L8_phase_diagram}
(a) Phase diagram of the correlated hopping model on the square lattice with half filling and background flux $\phi=\pi/2$, as a function of $J_\parallel/J_\perp$ and striped potential $\mu$, simulated on a cylinder of width $L=8$ sites. (b) Entanglement entropy and the variance of density in the perpendicular direction along a cut at fixed $J_\parallel/J_\perp = 0.5$. (c) Entanglement entorpy and correlation length along a cut at fixed $J_\parallel/J_\perp = 2$. (d) Density and nearest-neighbor correlations for the LL at $J_\parallel/J_\perp = 0.5, \mu=0$ and the CDW at $J_\parallel/J_\perp = 0.5, \mu=2.1$.}
\end{figure}

The iDMRG results shown in the main text for the square lattice with half filling and background flux $\phi=\pi/2$ are all obtained for system size $L=10$. While we have numerical evidence for the existence of the BIQH phase on systems of size $L=6,8,10$ [Fig.~\ref{fig:BIQH_LL}(a)], the phase diagram exhibits a marked difference for system sizes where $L/2$ is even or odd (compare Fig.~\ref{fig:L8_phase_diagram}(a) to Fig. 1 in the main text). This might be expected from the coupled chain picture: when $L/2$ is even, the system has a finite size gap which is not present in the odd case. In this section, we analyze the phase diagram for the even case by focusing on system size $L=8$. 

One major difference between the even and the odd case is the nature of the phase to the left of BIQH. The coupled wire analysis in the previous section predicts that an infinitesimal coupling $J_\parallel$ should drive the system into a BIQH state. However, the correlation length and entanglement entropy data in Fig.~\ref{fig:BIQH_LL}(c) suggests that the transition into the BIQH phase happens around $J_\parallel / J_\perp = 1.2$. Then a natural question arises: Is there an intermediate phase between the decoupled LL at $J_\parallel = 0$ and the BIQH phase? or should one think of this "intermediate phase" as one that is smoothly connected to the decoupled LL whose existence is due to the energy gap present at our finite size simulations?

Our numerical data supports the latter description for systems of even size, unlike the odd case. First, the correlation length connects smoothly to zero as $J_\parallel / J_\perp$ decreases from $1$ to $0$ [Fig.~\ref{fig:BIQH_LL}(c)], compared to the odd case where it diverges[Fig. 3(a) in the main text]. Furthermore, the large energy dispersion under flux insertion at $J_\parallel / J_\perp = 0.2$ [Fig.~\ref{fig:BIQH_LL}(b)] suggests that the state is gapless, which is consistent with the case of decoupled LLs. lastly, the correlation length data in Fig.~\ref{fig:BIQH_LL}(d) shows the critical point moving towards the decoupling limit as the inserted flux $\Phi$ is increased. This is also consistent with the decoupled LL scenario because the gap decreases with inserted flux and the results should more closely match what we expect to see in the thermodynamic limit.

As shown in Fig.~\ref{fig:L8_phase_diagram}, another major difference between the even and the odd case is the markedly distinct behavior of the system upon increasing $\mu$ for $1 \lesssim J_\parallel/J_\perp \lesssim 2$.
Rather than transitioning directly to the striped paramagnet phase from the BIQH phase, there appears to be an intermediate \emph{spontaneous} symmetry-breaking phase. 
This phase exhibits the same striped pattern as the paramagnet, but also spontaneously breaks symmetry in the perpendicular direction. 
The associated charge density wave (CDW) pattern is characterized by an order parameter measuring the variance of the density in the perpendicular direction [Fig.~\ref{fig:L8_phase_diagram}(b)]. Much like the shaded region in Fig. 1(c) in the main text, we caution that the region between the Luttinger Liquid phase and striped paramagnetic phase is difficult to converge in bond dimension due to its proximity to multiple phase transitions. It is possible, therefore, that the CDW does not represent the true ground state in the thermodynamic limit for all the green region in Fig.~\ref{fig:L8_phase_diagram}(a).
The presence of a strongly diverging correlation length as well as the smooth decrease in the order parameter suggests that the CDW-SPM transition is continuous.
Meanwhile, the  correlation length at the BIQH-CDW transition is only weakly enhanced as a function of the iDMRG bond dimension, and there is a sharp increase in the order parameter, consistent with the possibility of a first order phase transition; this is further supported by wave function overlap calculations.

\section{Experimental realization of the Floquet scheme}
The set up in Fig. 4 of the main text consists of four independent pairs of laser beams labeled as $\gamma=\{1,2,3,4\}$ (only two are shown for simplicity). Each pair consists of two vertically-polarized laser beams that are retro reflected and aligned along the primary axes of the lattice ($x$ and $y$ axis),

\begin{align}
\textbf{E}_\parallel^\gamma(\textbf{r},t) = &2 E_x \text{e}^{i\omega_\parallel^\gamma t} \cos (k x + \varphi_x^\gamma) \\
\textbf{E}_\perp^\gamma(\textbf{r},t) = &2 E_y \text{e}^{i[(\omega_\perp^\gamma + \Omega)t + \theta_0^\gamma]} \cos (k y + \varphi_y^\gamma) \\ \nonumber
+ &2 E_y \text{e}^{i[(\omega_\perp^\gamma - \Omega)t - \theta_0^\gamma]} \cos (k y + \varphi_y^\gamma) \\ \nonumber
= &4 E_y \text{e}^{i\omega_\perp^\gamma t}  \cos (\Omega t + \theta_0^\gamma) \cos (k y + \varphi_y^\gamma),
\end{align}

\noindent with the wave vectors chosen to be $|\mathbf{k}_\parallel|\simeq|\mathbf{k}_\perp|=k=\frac{\pi}{2a}$, where $a$ is the lattice constant. The laser beams along $y$ have two frequency components $\omega^{\gamma}_\perp \pm \Omega$ and the ones along $x$ a single component at $\omega^{\gamma}_\perp$, with $\omega^{\gamma}_\perp-\omega^{\gamma}_\parallel=\omega=\Delta_y/\hbar$ resonant with the potential energy difference between neighboring sites.
\noindent The total time-dependent potential that results from the interference of these beams consists of several terms $V_{\text{tot}} = V_{\text{cst}} + V_{2\Omega} + V_{\text{cross}}$. A constant part

\begin{equation} \label{eq:V_cst}
V^\gamma_{\text{cst}} = 4 E_x^2 \cos^2(kx+\varphi_x^\gamma) + 8 E_y^2 \cos^2(ky+\varphi_y^\gamma),
\end{equation}

\noindent and the second term

\begin{equation} \label{eq:V_Omega}
V^\gamma_{2\Omega} = 8 E_y^2 \cos(2\Omega t + 2 \theta_0^\gamma) \cos^2(k y+\varphi_y^\gamma),
\end{equation}

\noindent which generates a modulation at $2\Omega$. We will re-examine both terms at the end of this section. What we are interested in is the cross term

\begin{equation}
V^\gamma_{\text{cross}} = 16  E_xE_y \cos(\omega t) \cos(\Omega t + \theta_0^\gamma) \cos(kx+\varphi_x^\gamma) \cos(ky+\varphi_y^\gamma),
\end{equation}

In the high frequency limit $\omega \gg \Omega$, the term $\cos(\Omega t + \theta_0^\gamma)$ can be treated as constant and the dynamics are described by
\begin{align}
     H(t) = &-\sum_{m,n} \left(J_x \hat{a}^{\dagger}_{m+1,n} \hat{a}_{m,n} +J_y \hat{a}^{\dagger}_{m,n+1} \hat{a}_{m,n}+ \text{H.c.} \right) \\ \nonumber
     &+ \sum_{m,n} \left( n \Delta_y + \sum_{\gamma} V_0^\gamma \cos(\omega t) \cos(m\frac{\pi}{2} +\varphi_x^\gamma)  \cos(n\frac{\pi}{2} +\varphi_y^\gamma) \right) \hat{n}_{m,n},
\end{align}
where $m$ ($n$) is the lattice site index along $x$ ($y$), $V_0^\gamma = 16  E_xE_y \cos(\Omega t + \theta_0^\gamma)$. This Hamiltonian is periodic and can be approximated using a Floquet-Magnus expansion. The static term of the Hamiltonian contains diverging components proportional to $\hbar \omega = \Delta_y$. Therefore a transformation into the rotating frame is performed using the unitary operator
\begin{align}
    R_M(t) &= \exp \left[ i \sum_{m,n} \left( \frac{n \Delta_y t}{\hbar} + \sum_\gamma \frac{V_0^\gamma}{\hbar\omega} \sin(\omega t) \cos(m\frac{\pi}{2} +\varphi_x^\gamma)  \cos(n\frac{\pi}{2} +\varphi_y^\gamma) \right)  \hat{n}_{m,n} \right] \\ \nonumber
    &= \exp \left[ i \sum_{m,n} \chi_{m,n}(t) \hat{n}_{m,n} \right].
\end{align}
The transformed Hamiltonian can be written in the following form,
\begin{align}
    H_M(t) = &-\sum_{m,n} \left(J_x e^{i \eta^x_{m,n}(t)} \hat{a}^{\dagger}_{m+1,n} \hat{a}_{m,n} +J_y e^{i \eta^y_{m,n}(t)} \hat{a}^{\dagger}_{m,n+1} \hat{a}_{m,n}+ \text{H.c.} \right),
\end{align}
with $\eta^x_{m,n}(t) = \chi_{m+1,n}(t) - \chi_{m,n}(t)$ and $\eta^y_{m,n}(t) = \chi_{m,n+1}(t) - \chi_{m,n}(t)$ given by
\begin{align}
    \eta^x_{m,n}(t) &= -\eta^x_0 \sin(\omega t) \\
    \eta^y_{m,n}(t) &= -\eta^y_0 \sin(\omega t) + \frac{\Delta_y t}{\hbar},
\end{align}
where we have defined
\begin{align}
    \eta^x_0 &= \sum_\gamma  \frac{\sqrt{2} V_0^\gamma}{\hbar\omega} \cos(n\frac{\pi}{2} +\varphi_y^\gamma) \sin(m\frac{\pi}{2} + \frac{\pi}{4} +\varphi_x^\gamma) \\
    \eta^y_0 &= \sum_\gamma  \frac{\sqrt{2} V_0^\gamma}{\hbar\omega} \cos(m\frac{\pi}{2} +\varphi_x^\gamma) \sin(n\frac{\pi}{2} + \frac{\pi}{4} +\varphi_y^\gamma).
\end{align}
The lowest order of the time-independent Floquet Hamiltonian using the Magnus expansion is thus given by
\begin{align} \nonumber
    H_F &= \frac{1}{T} \int_0^T H_M(t) dt \\ \nonumber
    &= -\frac{1}{2\pi} \sum_{m,n} \left( J_x \hat{a}^{\dagger}_{m+1,n} \hat{a}_{m,n} \int_0^{2\pi} e^{-i \eta^x_0 \sin{\tau}} d\tau +J_y \hat{a}^{\dagger}_{m,n+1} \hat{a}_{m,n} \int_0^{2\pi} e^{i(\tau -\eta^y_0 \sin{\tau})} d\tau + \text{H.c.} \right) \\
    &= -\sum_{m,n} \left( J_x \mathcal{J}_0(\eta^x_0) \hat{a}^{\dagger}_{m+1,n} \hat{a}_{m,n} +J_y \mathcal{J}_1(\eta^y_0) \hat{a}^{\dagger}_{m,n+1} \hat{a}_{m,n} + \text{H.c.} \right)
\end{align}
where $\mathcal{J}_\nu(x) = \frac{1}{2\pi} \int_0^{2\pi} e^{i(\nu \tau - x \sin{\tau})} d\tau$ is the $\nu$th order Bessel function of the first kind. Using the expansion $\mathcal{J}_\nu(x) = \sum_n \frac{(-1)^n}{n! (n+\nu)!} (\frac{x}{2})^{2n+\nu}$, the Bessel functions can be approximated to the first order by $\mathcal{J}_0(x) \simeq 1$ and $\mathcal{J}_1(x) \simeq x/2$ for $x \ll 1$. Thus, in the limit $V_0 \ll \hbar \omega$, the effective Hamiltonian is given by
\begin{align}
    H_F \simeq -\sum_{m,n} ( J_x \hat{a}^{\dagger}_{m+1,n} \hat{a}_{m,n} + J_y^\text{eff} \hat{a}^{\dagger}_{m,n+1} \hat{a}_{m,n} + \text{H.c.} ),
    \label{eq:H_F}
\end{align}
with
\begin{equation} \label{eq:Jy_eff}
    J_y^\text{eff} = J_y \frac{\eta^y_0 }{2} =  J_y \frac{8 \sqrt{2} E_x E_y}{\Delta_y} \sum_\gamma  \cos(\Omega t + \theta_0^\gamma) \cos(m\frac{\pi}{2} +\varphi_x^\gamma) \sin(n\frac{\pi}{2} + \frac{\pi}{4} +\varphi_y^\gamma).
\end{equation}

\begin{figure}
\includegraphics[width=0.75\textwidth]{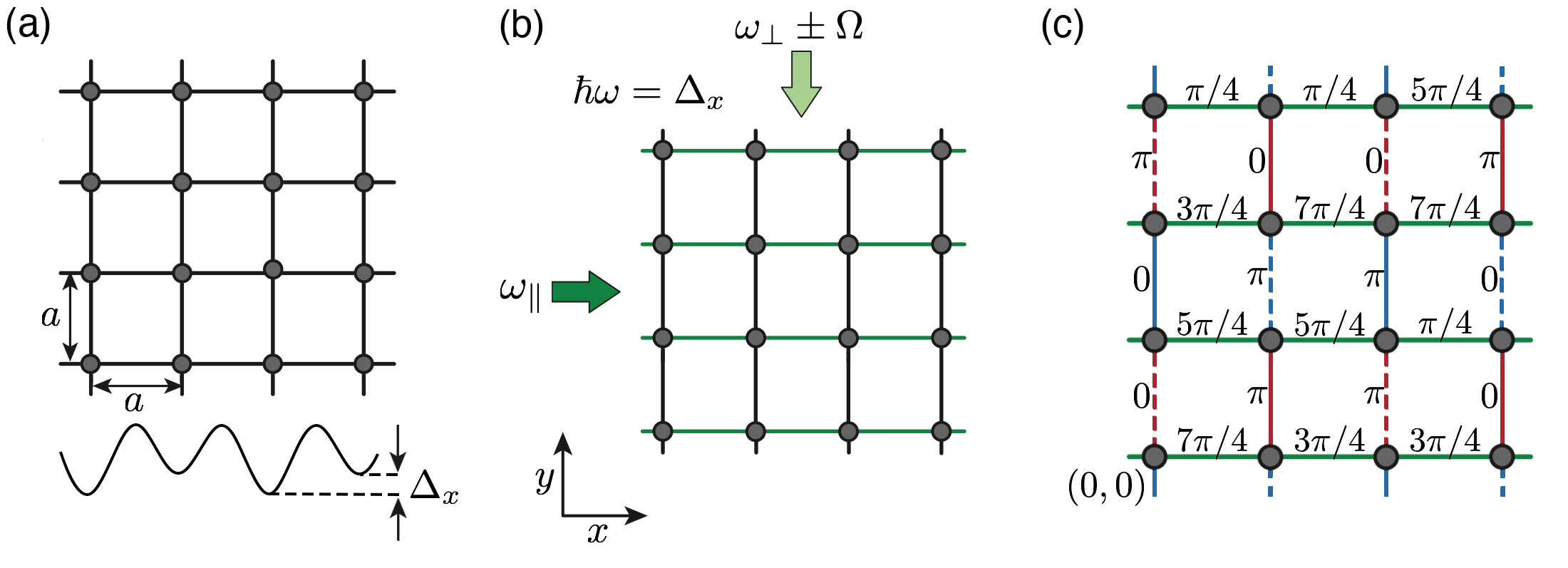} 
\caption{Second part of the experimental setup, where (a) the bosonic hopping $J_x$ is suppressed by an overlaid staggered potential $\Delta_x$, and (b) a pair of running-wave beams are used to restore resonant tunneling. (c) Phase distribution of the final effective Hamiltonian, realizing a homogenous background flux $\Phi=\pi/2$.
\label{fig:setup_2}}
\end{figure}

As desired, the effective coupling strength depends on $\cos(\Omega t + \theta_0^\gamma)$ and the spatial dependence $\cos(m\frac{\pi}{2} +\varphi_x^\gamma) \sin(n\frac{\pi}{2} + \frac{\pi}{4} +\varphi_y^\gamma)$ allows us to separately address the bonds with different values of $\theta_0^\gamma$. With the choice of $(\varphi_x^\gamma,\varphi_y^\gamma)=\{(0, \pi/4 ),(\pi/2, -\pi/4), (\pi, 3\pi/4), (\pi/2, \pi/4) \}$, we can address the red(dashed), blue(dashed), blue(solid), and red(solid) bonds respectively [Fig. \ref{fig:setup_2}(c)]. We then set $\theta_0^\gamma = \pi/2$ for dashed lines and $\theta_0^\gamma = 0$ for solid lines. Note that, under these parameters, the spatial dependence of $V^\gamma_\text{cst}$ and $V^\gamma_{2\Omega}$ for the combined setup gets cancelled and we obtain

\begin{align}
    \sum_\gamma{V^\gamma_\text{cst}} &= 8 E_x^2 + 16 E_y^2 \\
    \sum_\gamma{V^\gamma_{2\Omega}} &= 0.
\end{align}

The oscillation of the cos()sin() term in Eq. \ref{eq:Jy_eff} between $\pm 1$ results in a staggered background flux $\phi = 0, \pi$. In order to acquire a homogeneous flux as well as the amplitude modulation on the horizontal bonds, we repeat the same technique of laser-assisted tunneling in the perpendicular direction [Fig. \ref{fig:setup_2}(a,b)], by adding a staggered potential $\Delta_m = (-1)^m \Delta_x/2$ along the $x$ axis. A pair of running lasers with $\omega_\perp - \omega_\parallel = \omega = \Delta_x / \hbar$ is then used to restore the tunneling.

\begin{align}
\textbf{E}_\parallel(\textbf{r},t) = &E_x \text{e}^{i(\omega_\parallel t + k x)} \\
\textbf{E}_\perp(\textbf{r},t) = &E_y \text{e}^{i[(\omega_\perp + \Omega)t + k y + \theta_0]}
+ E_y \text{e}^{i[(\omega_\perp - \Omega)t + k y - \theta_0]} \\ \nonumber
= &2 E_y \text{e}^{i(\omega_\perp t + k y)}  \cos (\Omega t + \theta_0)
\end{align}
The resulting potential is
\begin{equation}
    V(\textbf{r},t) = \underbrace{E_x^2 + 2E_y^2}_{V_\text{cst}} + \underbrace{2E_y^2 \cos(2 \Omega t + 2\theta_0)}_{V_{2\Omega}} + \underbrace{4E_x E_y \cos(\Omega t + \theta_0) \cos(\omega t + kx - ky)}_{V_\text{cross}}.
\end{equation}
The first two terms don't have any spatial dependence, so again we are only interested in $V_\text{cross}$, leaving us with the following Hamiltonian.

\begin{align}
    \label{eq:H}
    H(t) = &-\sum_{m,n} \left(J_x \hat{a}^{\dagger}_{m+1,n} \hat{a}_{m,n} + J_y^\text{eff} \hat{a}^{\dagger}_{m,n+1} \hat{a}_{m,n}+ \text{H.c.} \right) \\ \nonumber
    &+ \sum_{m,n} \left( V_0 \sin(\omega t + \varphi_{m,n}) + (-1)^m \frac{\Delta_x}{2} \right) \hat{n}_{m,n},
\end{align}
where $V_0 = 4 E_x E_y \cos(\Omega t + \theta_0)$ and $\varphi_{m,n} = (m-n+1)\frac{\pi}{2}$.

Eq.~\ref{eq:H} has the exact same form as Eq. 5.22 of \cite{Saidelsburger2015artificial}. Following equations 5.23-5.29 of \cite{Saidelsburger2015artificial}, we obtain the final effective Hamiltonian
\begin{align}
    H_F = -\sum_{m,n} ( J_x^\text{eff} e^{i \Phi_{m,n}} \hat{a}^{\dagger}_{m+1,n} \hat{a}_{m,n} +J_y^\text{eff} \hat{a}^{\dagger}_{m,n+1} \hat{a}_{m,n} + \text{H.c.} ),
    \label{eq:H_F_final}
\end{align}
with
\begin{equation}
    J_x^\text{eff} = J_x \frac{1}{2} \frac{\sqrt{2} V_0}{\Delta_x} = J_x \frac{2 \sqrt{2} E_x E_y}{\Delta_x} \cos(\Omega t + \theta_0)
\end{equation}
\begin{equation}
    \Phi_{m,n} = \left\{
    \begin{array}{ll}
        -(\varphi_{m+1,n} + \varphi_{m,n})/2 & \mbox{for } m \mbox{ odd}\\
        +(\varphi_{m+1,n} + \varphi_{m,n})/2 + \pi & \mbox{for } m \mbox{ even}.
    \end{array}
    \right.
\end{equation}
As shown in Fig. \ref{fig:setup_2}(c), the combination of Peierls phases result in a homogeneous background flux $\Phi = \frac{\pi}{2}$ and we have constructed our desired Hamiltonian.

\section{Kibble-Zurek Treatment of the SPM-BIQH Transition}
As with every adiabatic preparation proposal, one needs to carefully consider whether the low-energy states of this exotic Hamiltonian can be prepared on an experimentally feasible time-scale. More specifically, the time-scale for adiabatic preparation of an exotic state tends to diverge as system size increases. However, the total experimental time is limited by decoherence, $t_\mathrm{dec}$. Therefore, the prepared state will only resemble a BIQH state on some limited length scale, $l_\mathrm{dec}$, set by $t_\mathrm{dec}$. In what follows, we will provide an estimate of $l_\mathrm{dec}$ assuming that we prepare the BIQH state using a quasi-adiabatic Kibble-Zurek~\cite{Schandran2012kibble} ramp of the striped chemical potential.

Let us fix $J_\perp = 1, J_\parallel = 2$ and start deep in the striped paramagnet phase, $\mu_i = 4$, where the ground state is well approximated by the easy-to-prepare fully striped state. We will start by assuming a linear ramp $\mu(t) = \mu_i - v t$, although more complex time-dependences can certainly lead to further improvements. 
Assuming the phase transition is continuous, as the striped potential is tuned closer to its critical value, $\mu_c \approx 2.0$, the correlation length and the relaxation time of the system diverge as 
\begin{align}
    \xi &= \xi_0 (\mu-\mu_c)^{-\nu} \label{corr_len_scaling} \\
    \tau &= \tau_0 (\mu-\mu_c)^{-z\nu} \label{tao_scaling}
\end{align}
with critical exponents $\nu \approx 0.5$ and $z = 1$. Note that these exponents have been taken  from the latest Monte-Carlo estimates~\cite{Sqin2017duality} for the theoretically-expected universality class of the phase transition between the BIQH and trivial state (i.e.~massless QED-3 with $N_f=2$ flavors of Dirac fermions) \cite{SGrover2013, SFuji2016}.

Due to this divergent relaxation time, there will always be a section of the Kibble-Zurek ramp sufficiently close to the critical point where the system cannot equilibrate fast enough and effectively enters a ``frozen" state.
This ``freeze-out'' time, $\bar{t}$, occurs when the time required for the ramp to reach the critical point is equal to the relaxation time:
\begin{equation}
    \bar{t} = \tau_0 [v \bar{t}]^{-z\nu} \Rightarrow \bar{t} = [\tau_0 v^{-z\nu}]^{1/(1+z\nu)}. 
\end{equation}
Then, the correlation length at this time provides a length scale, which characterizes the maximum size of the coherent domains that are formed during the adiabatic ramp:
\begin{equation}
    \bar{\xi} = \xi_0[v \bar{t}]^{-\nu} = \xi_0[\tau_0 v]^{-\nu/(1+z\nu)}.
\end{equation}
Note that the larger the ramp time (slower rate), the smaller the freeze-out time, and the larger the coherent domains will be. Let us assume that the maximum time which can be allotted to the adiabatic ramp is given by the experimental system's decoherence time-scale, $t_\mathrm{dec}$.
Then, the maximum length scale for coherent domains is given by
\begin{equation}
    l_\mathrm{dec} = \xi_0[\tau_0 (\mu_i - \mu_c)/t_\mathrm{dec}]^{-\nu/(1+z\nu)}. \label{l_dec}
\end{equation}
We can use our numerical data to estimate the constants of proportionality $\xi_0$ and $\tau_0$ from Eqs.~\ref{corr_len_scaling} and ~\ref{tao_scaling}. To do so, we extrapolate the data presented in Fig.~3(b) of the main text in bond dimension to obtain $\xi\approx 1.5$ at chemical potential $\mu=3$. This yields a corresponding estimate $\xi_0=1.5$ from Eqn.~\ref{corr_len_scaling}.
To estimate $\tau_0$, one can simply assume that the relaxation time is the inverse of the gap in units of $J_\perp$, i.e.~$\tau \approx 1/\Delta$. At a point inside the SPM phase with small correlation length $\xi \lesssim 1$, the gap can be approximated as the energy cost of moving one boson, i.e.~$\Delta = 2 \mu$. This regime overlaps with the critical-scaling regime $\xi \gtrsim 1$, in which Eqn.~\ref{tao_scaling} holds, when $\xi \sim 1$. 
From our numerics, this occurs when the chemical potential $\mu \approx 3.0$, which then implies $\tau_0 \approx 0.2$ from Eqn.~\ref{tao_scaling}.
Finally, although the decoherence time is strongly dependent on the specific  platform and driving protocol, current experiments suggest $t_\mathrm{dec} \approx 100 /J_\perp$ represents a reasonable time-scale~\cite{Sgross2017quantum, Sjepsen2020spin, Sye2022universal, Sbraun2024real}.
Using these estimates for the parameters, Eqn.~\ref{l_dec} yields a maximum correlation length (measured in units of the lattice spacing):
\begin{equation}
    l_\mathrm{dec} \simeq 9.5,
\end{equation}
which is an order of magnitude larger than the lattice spacing and thus, crucially, suggests the ability to adiabatically prepare a BIQH state which is significantly larger than a $\mathcal{O}(1)$ number of sites.
We note that in practice, we expect that one can  further  improve $l_\mathrm{dec}$ by considering a more optimized ramp profile than simply linear.

\end{document}